\documentclass[%
superscriptaddress,
twocolumn,
amsmath,amssymb,
prl,
floatfix,
]{revtex4-1}

\usepackage{graphicx}
\usepackage{bm}
\usepackage{footmisc}

\newcommand{\ANL}{HEP Division, Argonne National Laboratory, Lemont, IL}
\newcommand{\NIST}{National Institute of Standards and Technology, Boulder, CO}
\newcommand{\CU}{Department of Physics, University of Colorado, Boulder, CO}
\newcommand{\APM}{
    Key Laboratory of Atomic Frequency Standards,
    Innovation Academy for Precision Measurement Science and Technology, 
    Chinese Academy of Sciences, 
    Wuhan, China
}
\newcommand{\bra}[1]{\langle#1\rvert} 
\newcommand{\ket}[1]{\lvert#1\rangle} 

\newcommand{\da}{\downarrow}
\newcommand{\la}{\leftarrow}
\newcommand{\ra}{\rightarrow}

\newcommand{\ketda}{\ket{\downarrow}}
\newcommand{\ketua}{\ket{\uparrow}}

\newcommand{\sg}{\hat{\sigma}}
\newcommand{\Ham}{\hat{H}}
\newcommand{\am}{\hat{a}}
\newcommand{\amd}{\hat{a}^{\dagger}}

\newcommand{\Al}{{}^{27}\textrm{Al}^{+}}
\newcommand{\Mg}{{}^{25}\textrm{Mg}^{+}}
\newcommand{\sAl}{{}^1\textrm{S}_0}
\newcommand{\pAl}{{}^3\textrm{P}_1}
\newcommand{\clock}{{}^3\textrm{P}_0}

\begin{document}
\raggedbottom

\title{Scalable quantum logic spectroscopy}

\author{Kaifeng Cui}
\email{cuikaifeng@apm.ac.cn}
\affiliation{\NIST} 
\affiliation{\ANL}
\affiliation{\APM}  

\author{Jose Valencia}
\affiliation{\NIST}
\affiliation{\CU}

\author{Kevin T. Boyce}
\affiliation{\NIST}
\affiliation{\CU}

\author{Ethan R. Clements}
\affiliation{\NIST}
\affiliation{\CU}

\author{David R. Leibrandt}
\affiliation{\NIST}
\affiliation{\CU}

\author{David B. Hume}
\email{david.hume@nist.gov}
\affiliation{\NIST}

\date{\today}

\begin{abstract} 

In quantum logic spectroscopy (QLS), one species of trapped ion is used as a sensor to detect the state of an otherwise inaccessible ion species.  
This extends precision measurements to a broader class of atomic and molecular systems for applications like atomic clocks and tests of fundamental physics.
Here, we develop a new technique based on a Schr\"{o}dinger cat interferometer to address the problem of scaling QLS to larger ion numbers.  
We demonstrate the basic features of this method using various combinations of $\Mg$ logic ions and $\Al$ spectroscopy ions. We observe higher detection efficiency by increasing the number of $\Mg$ ions.
Applied to multiple $\Al$, this method will improve the stability of high-accuracy optical clocks and could enable Heisenberg-limited QLS.

\end{abstract}

\maketitle

Experiments on quantum systems face a common challenge of state detection, which requires amplifying tiny, quantum signals above the background noise.  
In the case of atomic systems, including trapped ions, the typical approach to state detection is to observe photons scattered from a particular quantum state~\cite{bergquist1986}.  
This approach works well on a limited number of atomic species that have a suitable internal structure; however, numerous other atomic species are compelling targets for specific applications but do not have suitable transitions for direct state detection.
For example, some species of molecular ions~\cite{wolf2016,chou2020}, highly-charged ions~\cite{micke2020}, and even antimatter particles~\cite{cornejo2021}, offer unique opportunities for testing fundamental physics~\cite{safronova2018}.  

Quantum logic spectroscopy (QLS) enables state detection of otherwise inaccessible ions by introducing a co-trapped logic ion (LI). The internal state of the spectroscopy ion (SI) is transferred via a shared mode of motion to the LI where it can be detected via photon scattering.
The first demonstration of QLS was with $^{27}$Al$^+$~\cite{schmidt2005, hume2007}, which now serves as a frequency reference in atomic clocks~\cite{beloy2021, hannig2019, cui2020,brewer2019}. 
Variations on the QLS technique have been developed with several aims: to demonstrate Hz-level precision spectroscopy for atomic clocks~\cite{beloy2021, hannig2019, cui2020,brewer2019},
to perform correlation spectroscopy~\cite{chou2011}, 
to work far off-resonance from an optical transition~\cite{hume2011, wolf2016}, 
or to operate in thermal motion~\cite{tan2015, kienzler2020}. 
So far QLS has only been performed on up to two SIs~\cite{chou2011} and the issue of scaling QLS techniques to larger ion numbers~\cite{schulte2016} is an open experimental question. 

\begin{figure}
    \centering
    \includegraphics[width=8.5cm]{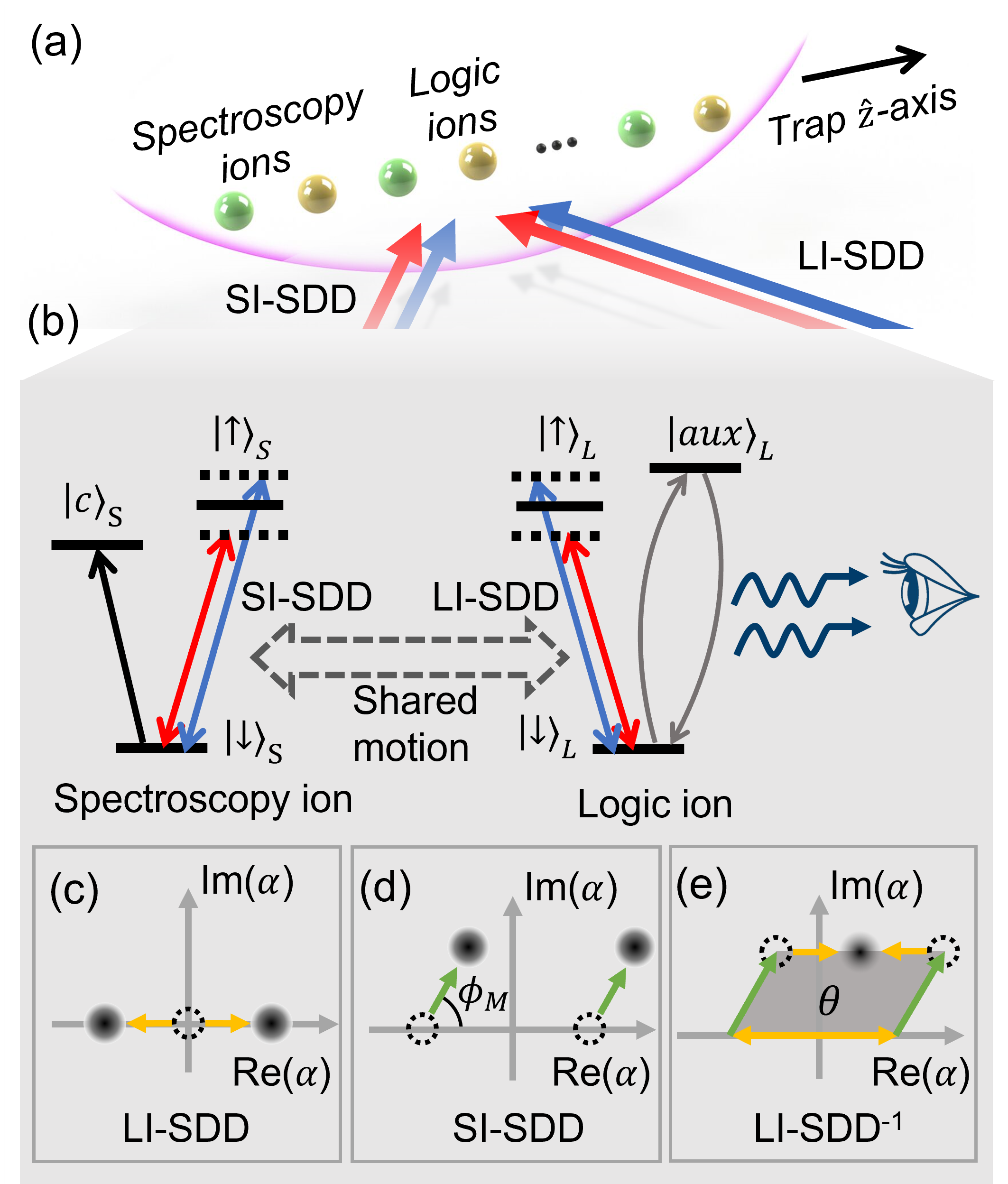}
    \caption{\label{fig:scheme}
    (a) An ensemble of spectroscopy ions (SIs, green dots) and logic ions (LIs, yellow dots) confined within a linear ion trap.
    The arrows indicate spin-dependent displacements(SDD) provided by two bichromatic laser fields.
    (b)The energy levels involved in these experiments.
    (c-e): Schr\"{o}dinger cat interferometry sequence in motional phase space.
    }
\end{figure}

In this paper, we propose and experimentally demonstrate a new method of QLS that can be scaled to larger ion numbers. 
Our protocol employs multiple LIs as independent sensors to detect a state-dependent driving force applied to the co-trapped SIs. 
This technique does not require ground state cooling or individual ion addressing, both of which become more difficult with larger ion ensembles~\cite{chen2020,feng2020}. 
We apply the technique experimentally to ensembles of up to 3 LIs and 3 SIs. 
By scaling the number of LIs, we show that technical noise in the detection process can be suppressed.  

Our protocol relies on a Schr\"{o}dinger cat state~\cite{monroe1996} of the LIs, which acts as an interferometric sensor for the state of the SIs (see Fig.~\ref{fig:scheme}). 
Several schemes have been explored to create the Schr\"{o}dinger cat states~\cite{monroe1996,milne2021,haljan2005, hempel2013}, including the $\hat{\sigma}_\phi$-type interaction~\cite{haljan2005, molmer1999} employed here. To produce this, a bichromatic laser field 
is applied to the LIs with frequency components near resonance with the motional sidebands at frequencies
$\omega_0 \pm \omega_M$, where $\omega_0$ is the qubit resonance frequency and $\omega_M$ is the frequency of a shared motional mode.
In the Lamb-Dicke limit, the dynamics of a single LI driven by the laser are described by the interaction Hamiltonian~\cite{wineland1998},
\begin{equation}
    \Ham = \frac{\hbar \eta \Omega_0}{2}\sg_\phi (\am e^{i\phi_M} +\amd e^{-i\phi_M}),
\end{equation}
where $2\pi\hbar$ is Planck's constant, $\Omega_0$ is the Rabi frequency, and $\eta$ is the Lamb-Dicke parameter,
describing the coupling strength between the laser field and the motional mode of the ions.
We use a rotated Pauli spin operator $\sg_\phi = e^{-i\phi_S}\sg_+ + e^{i\phi_S}\sg_-$ where $\sg_\pm$, $\am$, and $\amd$ are ladder operators of the spin mode and the motional mode, respectively.
The phases of the red ($\phi_r$) and blue ($\phi_b$) components of the laser field control the spin phase, $\phi_S=(\phi_b+\phi_r)/2$, and the motional phase, $\phi_M=(\phi_b-\phi_r)/2$.

\begin{figure}
    \centering
    \includegraphics[width=8.5cm]{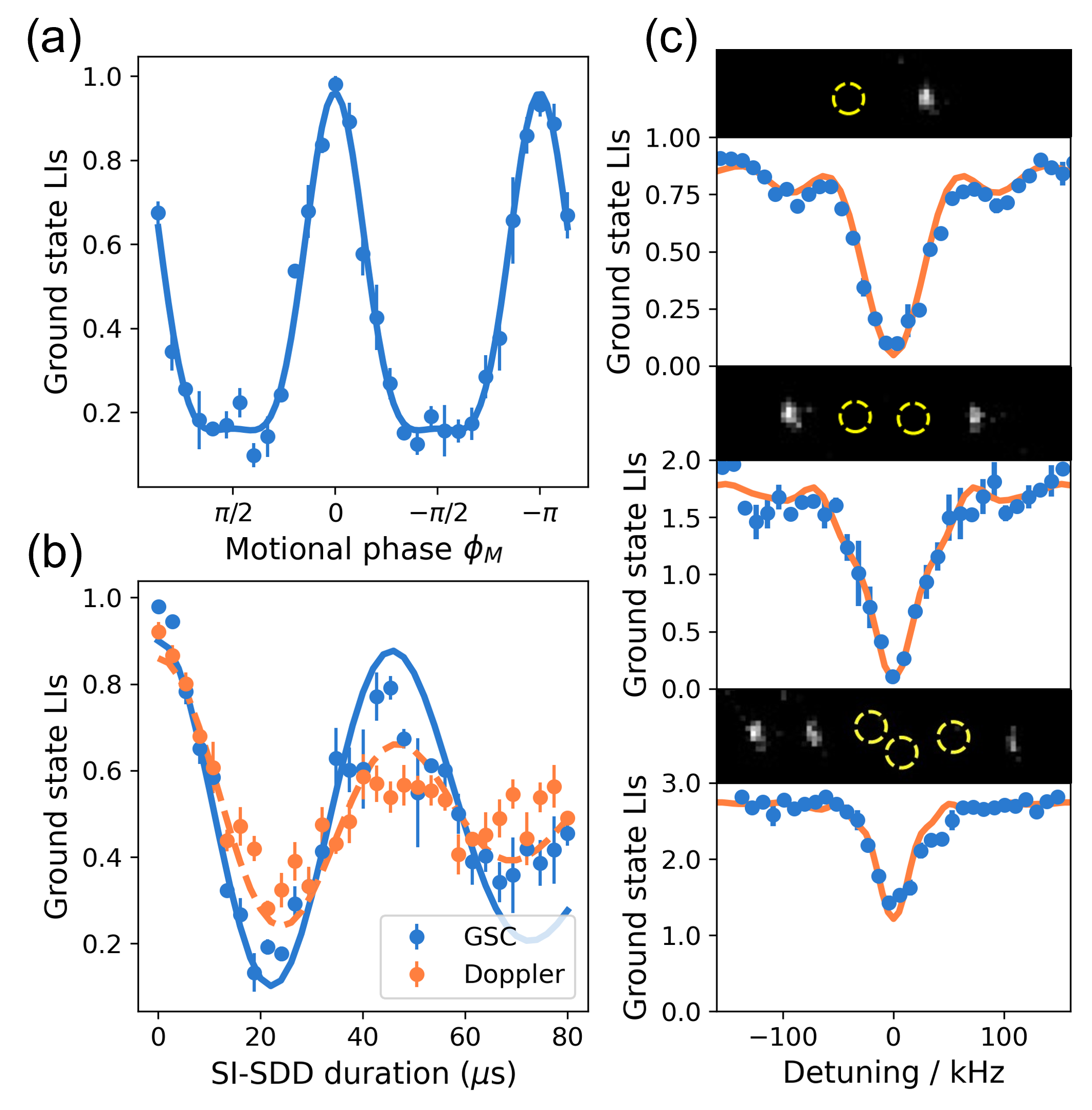}
    \caption{ \label{fig:scan_3p1}
        Quantum logic spectroscopy using a Schr\"{o}dinger cat interferometer.
        Using one $\Mg$ as a LI and one $\Al$ ion as a SI, the ground state probability of the LI ($P_{\da, L}$) is modulated by the geometric phase encoded in phase space when scanning the motional phase angle $\phi_M$.
        Experimental results (blue points) match the fit of Eq.~(\ref{measure}) (solid line). 
        (b) The duration of LI-SDD is scanned with ions cooled to the Doppler limit (orange points) and after sideband cooling (blue points).
        Lines are the results of numerical simulations without free parameters.
        (c) Spectroscopy of the $\sAl \ra \pAl$ transition of one, two, and three $\Al$ SIs using equal number of $\Mg$ LIs. 
        Insets on the top of each figure are fluorescence images of the $\Mg$ ions (bright spots).
        $\Al$ ion positions are marked in yellow circles based on theoretical calculations.
        The data (blue circles) are fit by numerical simulations (orange line). 
        Error bars represent one standard deviation of the LI quantum projection noise.
        }
\end{figure}

In the simplest case, 
consider a single LI prepared in a superposition state:
${\ket{\psi}_L} = {\ket{\da}_L} = {(\ket{\ra}_L + \ket{\la}_L)/\sqrt{2}}$,
where $\ketua_L$ and $\ketda_L$ are the energy eigenstates of the LI and $\ket{\la}_L$ and $\ket{\ra}_L$ are the eigenstates of $\hat{\sigma}_\phi$.
Applying the bichromatic laser field realizes a spin-dependent displacement on the LI (LI-SDD):
${\hat{\mathcal{U}} (t) = \hat{\mathcal{D}}(+\alpha)\ket{\ra}_L \bra{\ra}_L + \hat{\mathcal{D}}(-\alpha)\ket{\la}_L \bra{\la}_L}$,
where $\mathcal{D}$ is the motional phase space displacement amplitude and ${\alpha(t) = -i \eta \Omega_{0} t e^{-i\phi_M}/2}$. 
This is analogous to the first ``beam splitter'' of an interferometer, creating the Schr\"{o}dinger cat state shown in Fig.~1(c).
Likewise, the second beam splitter in the interferometer (LI-SDD$^{-1}$) can be produced by applying the same laser pulse, but shifting $\phi_M$ by $\pi$.  
If there is no additional displacement during the interferometer, this operation recombines the two motional components, recovering the initial state of the LI. 
However, displacements that occur between the two interferometer pulses generally produce a geometric phase, which can be detected in the final state of the LI.  Here, we consider a state-dependent displacement (SI-SDD) produced by another bichromatic laser field applied to a SI coupled to the LI through their collective motion at frequency $\omega_M$.
For example, if the SI is prepared in the state: $\ket{\psi}_S = {(\ketda_S + \ketua_S)/\sqrt{2}} = \ket{\ra}_S$
it undergoes a displacement $\hat{\mathcal{D}}(\beta)$ (marked as SI-SDD in Fig.~1(d)), which produces a geometric phase $\theta = 2 \alpha \beta \sin\phi_M$, rotating the state of the LI by $e^{-2i\theta \hat{\sigma}_\phi}$.
Measurement of the LI population $P_{\da, L}$ after the second beamsplitter gives:
\begin{equation}
    \label{measure}
    P_{\da, L} = [1+\cos(4 \alpha \beta \sin\phi_M)]/2.    
\end{equation}

\begin{figure}
    \centering
    \includegraphics[width=8.5cm]{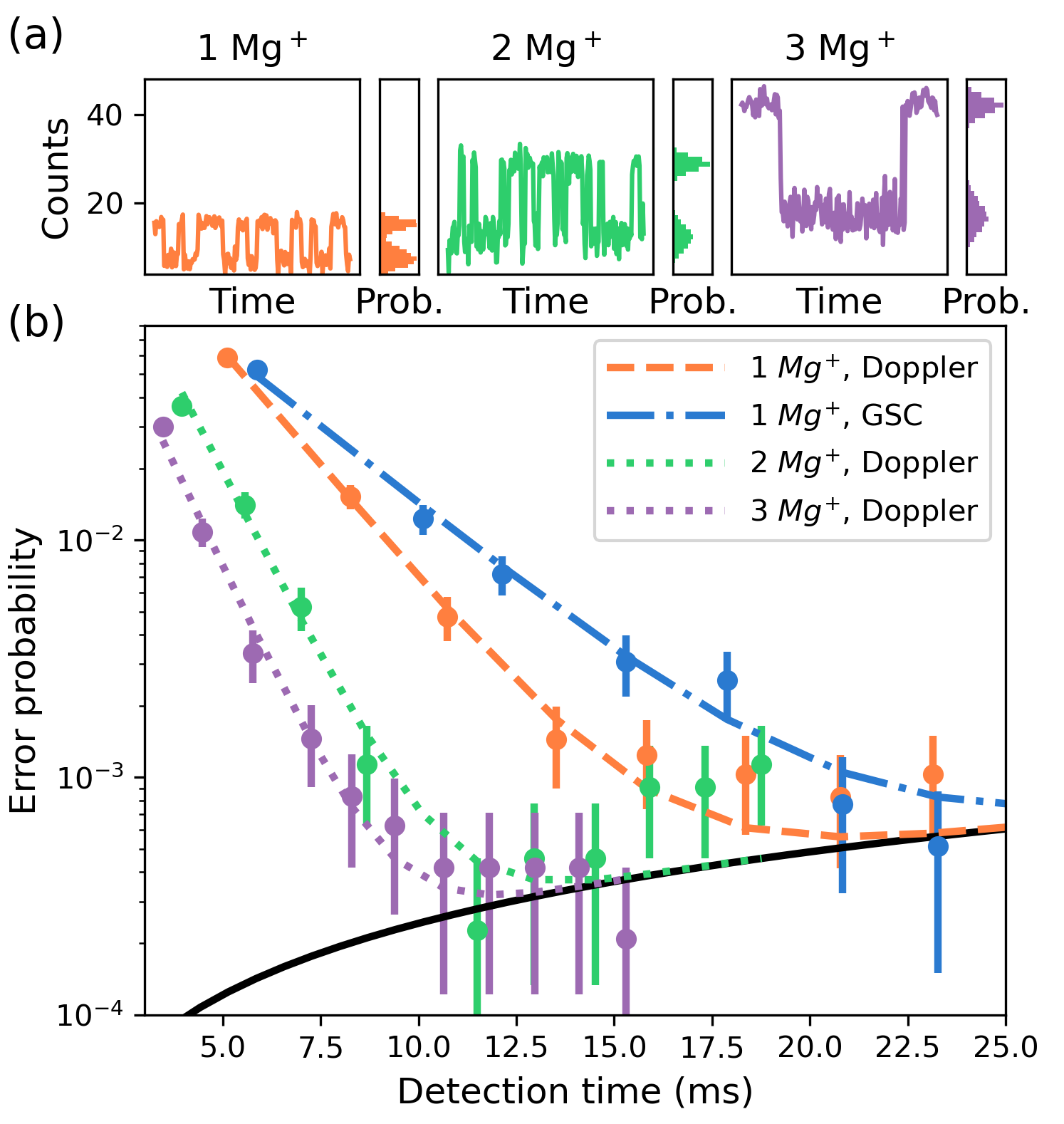}
    \caption{\label{fig:result_qnd}
    Scaling the number of LIs to improve the detection efficiency of a single SI.
    (a) Transitions between $\ketda_S$ and $\ket{c}_S$ control the output of the interferometer resulting in quantum jumps in the fluorescence of the LIs.
    Each of the data points is an average of 20 measurements
    where a single measurement takes 1.8 ms.
    The histograms on the right show the full distribution of these data (1500 data points, Prob.: probability density).
    (b) Observed error rate estimate comes from the comparison of two consecutive detection sequences during the detection.
    Error bars are symmetric, based on the standard variance of a Poisson distribution.
    The black solid line represents the lifetime-limited error probability 
    due to the spontaneous decay of the $\ket{c}_S$ state. 
    }
\end{figure}

The parameter $\beta$ contains information about the interaction between the SI and the SI-SDD beams, which is detected interferometrically by the LI.  
We use this in two distinct ways.  
First, we perform spectroscopy directly on the $\ketda_S\leftrightarrow\ketua_S$ transition, where the phase, duration and detuing of the SI-SDD pulse itself modulates $\beta$ (Fig.~\ref{fig:scan_3p1}).  
Second, we detect ``clock’’ transitions from $\ketda_S$ to the long-lived clock state ($\ket{c}_S$) using the fact that only the ions in state $\ketda_S$ interact with the SI-SDD pulse.  
This allows for spectroscopy on a narrow clock transition and is analogous to the electron-shelving technique used in conventional fluorescence measurements (Fig.~\ref{fig:result_qnd}).

Both protocols can be scaled to $N_L$ LIs and $N_S$ SIs.  
Assuming that all ions have nearly equal mode amplitudes and feel equal driving forces, 
all the LIs can be treated as independent sensors.
The signal observed by the LIs increases linearly with $N_L$~\footnotemark[100].  
In scaling $N_S$, the force experienced by the ions during the SI-SDD pulse increases linearly with the number SIs in the state $\ketda_S$.  
By appropriate choice of parameters $\alpha$ and $\phi_M$ this provides a means to count the number of ions remaining in $\ketda_S$ after a clock pulse on multiple SIs.  

We demonstrate this interferometer using $\Mg$ as the LI and $\Al$ as the SI confined in a linear Paul trap~\cite{chen2017,brewer2019}.
The trap frequencies are approximately  $(\omega_x, \omega_y, \omega_z) = 2\pi \times (6.7, 6.3, 2.5)$ MHz for a single $\Mg$.
The qubit states of the $\Mg$ ions are encoded in 
${\ketda_L \equiv \ket{^2S_{1/2}, F = 3, m_F = -3}}$ and ${\ketua_L \equiv \ket{^2S_{1/2}, F = 2, m_F = -2}}$.
Doppler cooling and state detection of $\Mg$ relies on resonance fluorescence
from the ${{\ketda_L} \leftrightarrow \ket{^2P_{3/2}, F = 4, m_F = -4}}$ cycling transition driven by a circularly-polarized 280.4 nm laser beam.

A pair of perpendicular 279.6 nm laser beams, referred to here as red Raman (RR) and blue Raman (BR) respectively, 
with wavevector difference $\Delta \bm{k}$ along the trap $\bm{z}$-axis are used to generate Raman sideband pulses~\cite{chen2017}, and the LI-SDD.
The BR beam consists of two frequency components, which are separated by $2\omega_M$.
For the $\Al$ ions, the qubit states are encoded in 
${\ketda_S} \equiv \ket{\sAl, m_F=-5/2}$ and 
${\ketua_S} \equiv \ket{\pAl, m_F=-7/2}$.
A circular-polarized, bichromatic 266.9 nm beam line is applied at a 45$^{\circ}$ angle with the trap $z$-axis for the qubit manipulation and the SI-SDD. 
Both the 266.9 nm and the 279.6 nm laser beams are intensity stabilized using photodiodes before entering the trap to generate carrier Rabi rates of approximately 300 kHz. 
When trapping one $\Mg$ ion and one $\Al$ ion, 
the Lamb-Dicke parameters of the 2.5-MHz center-of-mass~(COM) mode along the $\bm{z}$-axis are $\eta_L=0.18$ for the LI and $\eta_S=0.10$ for the SI, respectively.
We use this mode to drive both the LI-SDD and the SI-SDD as the mode amplitudes $z_0$ are nearly the same for ions in different positions.
In addition, the motional phases of both SDDs need to be equal for all the ions and controllable between species, which can be satisfied using a bichromatic laser field~\footnotemark[100].
When scaling to multiple number of ions,  
those Lamb-Dicke parameters decrease since the ground-state wavefunction size $z_0 \propto (M \omega_M)^{-1/2}$, where $M$ is the total mass of the ion chain.

We first prepare a pair of $\Mg$ and $\Al$ to demonstrate features of the interferometer.
The $\Mg$ ion is optically pumped to the $\ketda_L$ state, while the $\Al$ ion is rotated to 
$\ket{\ra}_S= ({\ketda_S} + {\ketua_S})/2$ using a $\pi/2$-carrier pulse. 
In order to control the geometric phase enclosed in the interferometer, it is necessary
to maintain the relative phases between the LI-SDD on the $\Mg$ ions and SI-SDD on the $\Al$ ions.
To accomplish this, we produce the red and blue tones for the two bichromatic laser beams by mixing radio-frequency signals that are used to drive two acousto-optic modulators (AOMs) from a single source~\footnotemark[100].
The long-term phase coherence can be observed between these two pairs of laser beams
by scanning their relative phase $\phi_M$ (Fig.~\ref{fig:scan_3p1}(a)).
In this experiment, we have calibrated $4\alpha\beta=\pi$ and the solid line is a fit based on Eq.~(3).  In the following experiments we set $\phi_M=\pi/2$ to maximize the geometric phase.

By scanning the duration of the SI-SDD (Fig.~\ref{fig:scan_3p1}) we observe how the detection signal varies as a function of $\beta$.
We include experimental results when the ions are cooled close to the Doppler limit and with 1.25 ms of additional sideband cooling (SBC).
Both cases are affected by higher-order terms in the Hamiltonian beyond the Lamb-Dicke limit,
which appears as a loss in contrast of the detection signal as a function of the SI-SDD pulse time.
Due to higher temperature, this effect is more significant for the Doppler-cooled case.

We use this interferometer to perform spectroscopy of the $\sAl \rightarrow \pAl$ transition of up to three $\Al$ ions co-trapped with the same number of $\Mg$ ions.
The duration of both LI-SDD and SI-SDD pulses are calibrated on resonance to make the geometric phase enclosed within the interferometer $2N_{S}\alpha\beta=\pi/2$. 
The detuning between the 266.9 nm laser beam and the $\sAl \rightarrow \pAl$ transition ($\delta_s$) is scanned over a range of 300 kHz.
Additional SBC pulses are applied for all the cases to suppress the coupling outiside the Lamb-Dicke regime. 
We note that when $\delta_s \neq 0$, the $\Al$ ions and the collective motion undergo a complicated evolution,
resulting in a complex line shape shown in the simulation results~\footnotemark[100].
For the case of 6 ions, the ion chain has formed a zig-zag geometry, but we still observe a resonant response in the interferometer, although with reduced contrast.

Now we introduce the SI clock state ($\ket{c}_S$) encoded in the $\ket{\clock, m_F=-5/2}$ state which has a lifetime of approximately 20.6 s~\cite{rosenband2007}.  Precision measurement of the $\ketda_S\leftrightarrow\ket{c}_S$ transition is the basis for Al$^+$ optical clocks.  
We use a single $\Al$ ion and vary the number of $\Mg$ ions from 1 to 3. The $\Al$ ion is driven periodically by a weak 267.4 nm laser that is close to resonance with 
the clock transition resulting in infrequent state changes.
These ``quantum jumps'' are clearly observed (Fig. 3(a)) since the SI-SDD, and hence the fluorescence of the $\Mg$ ions is gated by the state of the $\Al$ ion.

We select the number of measurement repetitions via an adaptive Bayesian process~\cite{hume2007}. 
The detection error probability is determined by comparing the results of 
two consecutive detection sequences~(Fig.~\ref{fig:result_qnd}(b)),  counting one detection error if they disagree.  
This assumes that the probability of two consecutive detection errors can be ignored, which is true for small, uncorrelated error probabilities.  We expect this to be valid for shorter detection times (the initial slopes in  Fig.~3(b)), but errors due to spontaneous decay at longer detection times will violate the assumption of uncorrelated errors~\cite{erickson2021}.   
We observe that increasing the number of $\Mg$ ions increases the measurement efficiency.
In addition to the improved signal-to-noise ratio, given the same confinement conditions,
the Lamb-Dicke parameters are also reduced with more LIs,
improving the contrast by suppressing imperfections due to higher-order processes.

\begin{figure}
\centering
\includegraphics[width=6.5cm]{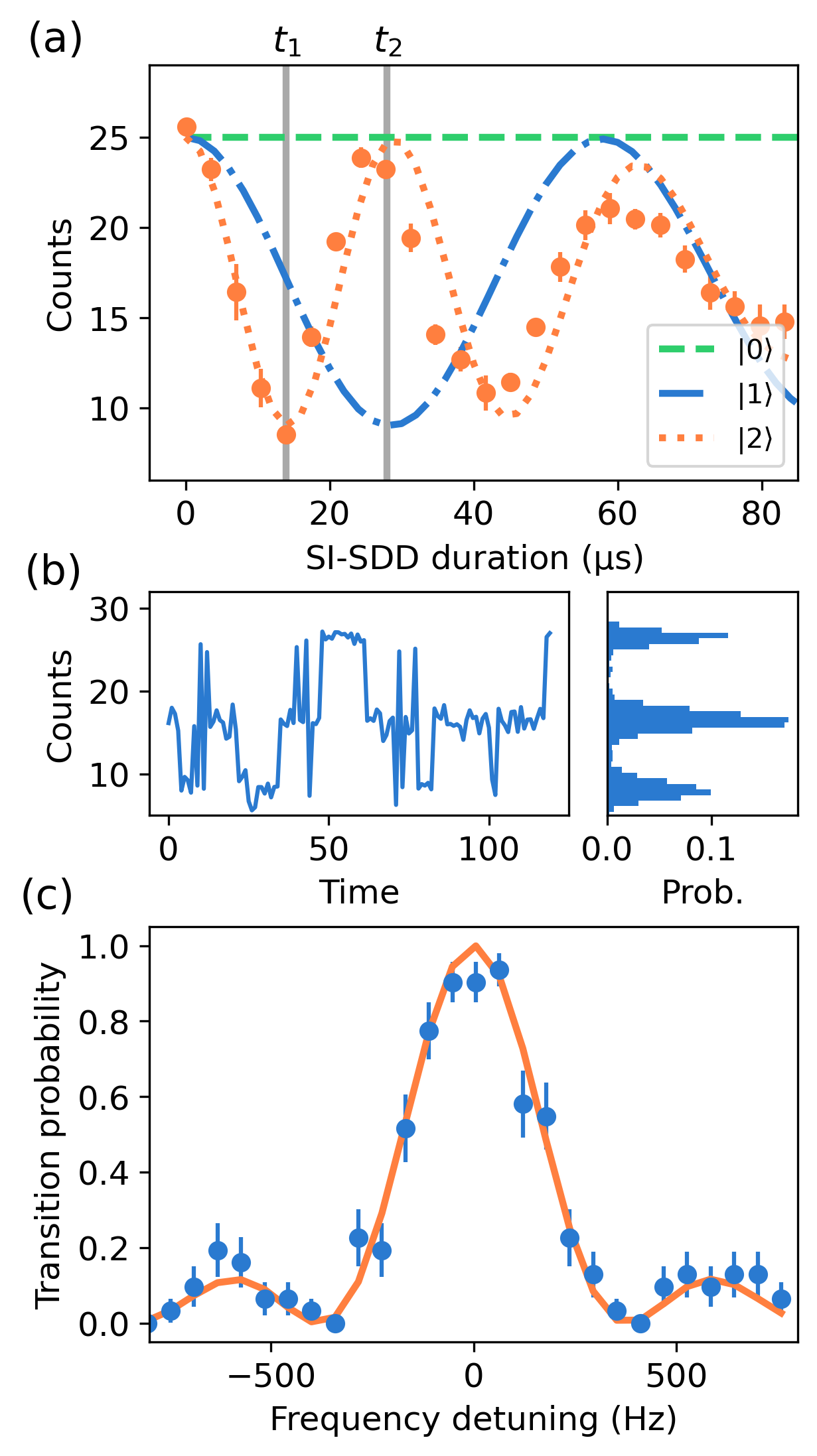}
\caption{\label{fig:result_2Al}
    (a) $\Mg{}$ fluorescence when scanning the duration of the SI-SDD pulse.
    The observed fluorescence for $N_S = 2$ is plotted along with the expected signal for $N_S$ = 1 and 0.
    All lines come from numerical simulations using measured experimental parameters.
    (b) Three-level ``quantum jumps'' using two $\Mg$ and two $\Al$ ions.
    Each data point is an average of 100 measurements.
    (c) Quantum logic spectroscopy of the $\sAl \leftrightarrow \clock$ transition of two $\Al$ SIs.
    All experiments in this figure are done after 1.25 ms sideband cooling. Error bars represent one standard deviation of the mean.
}
\end{figure}

In Fig.~\ref{fig:result_qnd}(b), we also compare the efficiency of detection at the Doppler limit versus after sideband cooling the z-COM mode to near the motional ground state.
Although the single-shot fidelity is lower with only Doppler cooling, 
the 1.25 ms additional duration of the sideband cooling sequence makes it less efficient.


We demonstrate the detection protocol with two $\Al$ and two $\Mg$ ions ($N_S = N_L = 2$).
We introduce the number states $\ket{N_{S,\downarrow}}$ to represent the number of SIs remaining in state $\ketda_{S}$,
where $\ket{N_{S,\downarrow}}\in\{\ket{0},\ket{1}, \ket{2}\}$.
In Fig.~\ref{fig:result_2Al}(a), we show expected fluorescence levels for each of these three cases as a function of the SI-SDD duration.  Choosing the SI-SDD duration to be $t_1 = 14$ $\rm{\mu s}$ results in an interferometer phase of $\theta = \pi/2$ for the case $N_{S,\downarrow} = 2$, corresponding to both logic ions flipping from bright to dark.  
However, for $N_{S,\downarrow} = 1$ there is a 50~\% transition probability for both LIs, and for $N_{S,\downarrow} = 0$ both LIs remain bright.  
In Fig.~\ref{fig:result_2Al}(b) we prepare these states probabilistically by scanning the frequency of the 267.4 nm laser over the resonance frequency of the $\ketda_S\rightarrow\ket{c}_S$ transition.  
The final determination of the number state uses the same adaptive Bayesian process described earlier, yielding the spectroscopy of the $\sAl \rightarrow \clock$ transition as shown in  Fig.~\ref{fig:result_2Al}(c).
We calculate projection noise limits as a function of both $N_L$ and $N_S$ in~\footnotemark[100].  
Ideally, technical noise $\sigma_{N_S}$ in determination of $N_{S,\downarrow}$ is negligible compared to the fundamental quantum projection noise limit.

The choice of $\alpha$ and $\beta$ allow the efficiency of the detection sequence to be optimized and adapted to different measurement bases.
For example, if we choose $\alpha\beta = N_S \pi/2$ ($t=t_2$ in Fig.~4~(a)), even-parity SI states will result in all LIs bright, whereas odd-parity SI states will result in all LIs dark. 
This provides an efficient means for performing QLS at the Heisenberg limit~\cite{bollinger1996}.

In summary, we have demonstrated a method of performing QLS based on the Schr\"{o}dinger cat interferometer that allows for scaling both the number of spectroscopy ions and logic ions.  This technique operates with ions in thermal motion and is insensitive to the position of the ions in the array.  Technical  improvements to the current experiment including higher laser power to address all ions equally and improved background gas pressure will allow this protocol to be scaled to longer ion chains. 
As shown in the supplemental material~\footnotemark[100], 
the projection noise in a single detection cycle using this protocol depends on the number of logic ions in addition to the number of spectroscopy ions.
In the future, this technique could allow quantum logic spectroscopy in even larger ion ensembles, where some of the same capabilities used here have already been demonstrated~\cite{gilmore2017,affolter2020}. 

\section*{Acknowledgments}

We thank D. Barberena for useful discussion, 
J.J. Bollinger, A.M. Rey and K. Beloy for their careful reading and feedback on this manuscript. 
This work was supported by the National Institute of Standards and Technology, the National Science Foundation Q-SEnSE Quantum Leap Challenge Institute (Grant Number 2016244), and the Office of Naval Research (Grant Number N00014-18-1-2634). K.C. was supported by a DOE Office of Science HEP QuantISED award.

\footnotetext[100]{See supplemental material}
\nocite{arecchi1972}
\nocite{wineland1998}

\bibliography{multi_ions}

\begin{thebibliography}{33}%
\makeatletter
\providecommand \@ifxundefined [1]{%
 \@ifx{#1\undefined}
}%
\providecommand \@ifnum [1]{%
 \ifnum #1\expandafter \@firstoftwo
 \else \expandafter \@secondoftwo
 \fi
}%
\providecommand \@ifx [1]{%
 \ifx #1\expandafter \@firstoftwo
 \else \expandafter \@secondoftwo
 \fi
}%
\providecommand \natexlab [1]{#1}%
\providecommand \enquote  [1]{``#1''}%
\providecommand \bibnamefont  [1]{#1}%
\providecommand \bibfnamefont [1]{#1}%
\providecommand \citenamefont [1]{#1}%
\providecommand \href@noop [0]{\@secondoftwo}%
\providecommand \href [0]{\begingroup \@sanitize@url \@href}%
\providecommand \@href[1]{\@@startlink{#1}\@@href}%
\providecommand \@@href[1]{\endgroup#1\@@endlink}%
\providecommand \@sanitize@url [0]{\catcode `\\12\catcode `\$12\catcode
  `\&12\catcode `\#12\catcode `\^12\catcode `\_12\catcode `\%12\relax}%
\providecommand \@@startlink[1]{}%
\providecommand \@@endlink[0]{}%
\providecommand \url  [0]{\begingroup\@sanitize@url \@url }%
\providecommand \@url [1]{\endgroup\@href {#1}{\urlprefix }}%
\providecommand \urlprefix  [0]{URL }%
\providecommand \Eprint [0]{\href }%
\providecommand \doibase [0]{http://dx.doi.org/}%
\providecommand \selectlanguage [0]{\@gobble}%
\providecommand \bibinfo  [0]{\@secondoftwo}%
\providecommand \bibfield  [0]{\@secondoftwo}%
\providecommand \translation [1]{[#1]}%
\providecommand \BibitemOpen [0]{}%
\providecommand \bibitemStop [0]{}%
\providecommand \bibitemNoStop [0]{.\EOS\space}%
\providecommand \EOS [0]{\spacefactor3000\relax}%
\providecommand \BibitemShut  [1]{\csname bibitem#1\endcsname}%
\let\auto@bib@innerbib\@empty
\bibitem [{\citenamefont {Bergquist}\ \emph {et~al.}(1986)\citenamefont
  {Bergquist}, \citenamefont {Hulet}, \citenamefont {Itano},\ and\
  \citenamefont {Wineland}}]{bergquist1986}%
  \BibitemOpen
  \bibfield  {author} {\bibinfo {author} {\bibfnamefont {J.~C.}\ \bibnamefont
  {Bergquist}}, \bibinfo {author} {\bibfnamefont {R.~G.}\ \bibnamefont
  {Hulet}}, \bibinfo {author} {\bibfnamefont {W.~M.}\ \bibnamefont {Itano}}, \
  and\ \bibinfo {author} {\bibfnamefont {D.~J.}\ \bibnamefont {Wineland}},\
  }\href {\doibase 10.1103/PhysRevLett.57.1699} {\bibfield  {journal} {\bibinfo
   {journal} {Phys. Rev. Lett.}\ }\textbf {\bibinfo {volume} {57}},\ \bibinfo
  {pages} {1699} (\bibinfo {year} {1986})}\BibitemShut {NoStop}%
\bibitem [{\citenamefont {Wolf}\ \emph {et~al.}(2016)\citenamefont {Wolf},
  \citenamefont {Wan}, \citenamefont {Heip}, \citenamefont {Gebert},
  \citenamefont {Shi},\ and\ \citenamefont {Schmidt}}]{wolf2016}%
  \BibitemOpen
  \bibfield  {author} {\bibinfo {author} {\bibfnamefont {F.}~\bibnamefont
  {Wolf}}, \bibinfo {author} {\bibfnamefont {Y.}~\bibnamefont {Wan}}, \bibinfo
  {author} {\bibfnamefont {J.~C.}\ \bibnamefont {Heip}}, \bibinfo {author}
  {\bibfnamefont {F.}~\bibnamefont {Gebert}}, \bibinfo {author} {\bibfnamefont
  {C.}~\bibnamefont {Shi}}, \ and\ \bibinfo {author} {\bibfnamefont {P.~O.}\
  \bibnamefont {Schmidt}},\ }\href {\doibase 10.1038/nature16513} {\bibfield
  {journal} {\bibinfo  {journal} {Nature}\ }\textbf {\bibinfo {volume} {530}},\
  \bibinfo {pages} {457} (\bibinfo {year} {2016})}\BibitemShut {NoStop}%
\bibitem [{\citenamefont {Chou}\ \emph {et~al.}(2020)\citenamefont {Chou},
  \citenamefont {Collopy}, \citenamefont {Kurz}, \citenamefont {Lin},
  \citenamefont {Harding}, \citenamefont {Plessow}, \citenamefont {Fortier},
  \citenamefont {Diddams}, \citenamefont {Leibfried},\ and\ \citenamefont
  {Leibrandt}}]{chou2020}%
  \BibitemOpen
  \bibfield  {author} {\bibinfo {author} {\bibfnamefont {C.~W.}\ \bibnamefont
  {Chou}}, \bibinfo {author} {\bibfnamefont {A.~L.}\ \bibnamefont {Collopy}},
  \bibinfo {author} {\bibfnamefont {C.}~\bibnamefont {Kurz}}, \bibinfo {author}
  {\bibfnamefont {Y.}~\bibnamefont {Lin}}, \bibinfo {author} {\bibfnamefont
  {M.~E.}\ \bibnamefont {Harding}}, \bibinfo {author} {\bibfnamefont {P.~N.}\
  \bibnamefont {Plessow}}, \bibinfo {author} {\bibfnamefont {T.}~\bibnamefont
  {Fortier}}, \bibinfo {author} {\bibfnamefont {S.}~\bibnamefont {Diddams}},
  \bibinfo {author} {\bibfnamefont {D.}~\bibnamefont {Leibfried}}, \ and\
  \bibinfo {author} {\bibfnamefont {D.~R.}\ \bibnamefont {Leibrandt}},\ }\href
  {\doibase 10.1126/science.aba3628} {\bibfield  {journal} {\bibinfo  {journal}
  {Science}\ }\textbf {\bibinfo {volume} {367}},\ \bibinfo {pages} {1458}
  (\bibinfo {year} {2020})}\BibitemShut {NoStop}%
\bibitem [{\citenamefont {Micke}\ \emph {et~al.}(2020)\citenamefont {Micke},
  \citenamefont {Leopold}, \citenamefont {King}, \citenamefont {Benkler},
  \citenamefont {Spie{\ss}}, \citenamefont {Schm{\"o}ger}, \citenamefont
  {Schwarz}, \citenamefont {{Crespo L{\'o}pez-Urrutia}},\ and\ \citenamefont
  {Schmidt}}]{micke2020}%
  \BibitemOpen
  \bibfield  {author} {\bibinfo {author} {\bibfnamefont {P.}~\bibnamefont
  {Micke}}, \bibinfo {author} {\bibfnamefont {T.}~\bibnamefont {Leopold}},
  \bibinfo {author} {\bibfnamefont {S.~A.}\ \bibnamefont {King}}, \bibinfo
  {author} {\bibfnamefont {E.}~\bibnamefont {Benkler}}, \bibinfo {author}
  {\bibfnamefont {L.~J.}\ \bibnamefont {Spie{\ss}}}, \bibinfo {author}
  {\bibfnamefont {L.}~\bibnamefont {Schm{\"o}ger}}, \bibinfo {author}
  {\bibfnamefont {M.}~\bibnamefont {Schwarz}}, \bibinfo {author} {\bibfnamefont
  {J.~R.}\ \bibnamefont {{Crespo L{\'o}pez-Urrutia}}}, \ and\ \bibinfo {author}
  {\bibfnamefont {P.~O.}\ \bibnamefont {Schmidt}},\ }\href {\doibase
  10.1038/s41586-020-1959-8} {\bibfield  {journal} {\bibinfo  {journal}
  {Nature}\ }\textbf {\bibinfo {volume} {578}},\ \bibinfo {pages} {60}
  (\bibinfo {year} {2020})}\BibitemShut {NoStop}%
\bibitem [{\citenamefont {Cornejo}\ \emph {et~al.}(2021)\citenamefont
  {Cornejo}, \citenamefont {Lehnert}, \citenamefont {Niemann}, \citenamefont
  {Mielke}, \citenamefont {Meiners}, \citenamefont {{Bautista-Salvador}},
  \citenamefont {Schulte}, \citenamefont {Nitzschke}, \citenamefont {Borchert},
  \citenamefont {Hammerer}, \citenamefont {Ulmer},\ and\ \citenamefont
  {Ospelkaus}}]{cornejo2021}%
  \BibitemOpen
  \bibfield  {author} {\bibinfo {author} {\bibfnamefont {J.~M.}\ \bibnamefont
  {Cornejo}}, \bibinfo {author} {\bibfnamefont {R.}~\bibnamefont {Lehnert}},
  \bibinfo {author} {\bibfnamefont {M.}~\bibnamefont {Niemann}}, \bibinfo
  {author} {\bibfnamefont {J.}~\bibnamefont {Mielke}}, \bibinfo {author}
  {\bibfnamefont {T.}~\bibnamefont {Meiners}}, \bibinfo {author} {\bibfnamefont
  {A.}~\bibnamefont {{Bautista-Salvador}}}, \bibinfo {author} {\bibfnamefont
  {M.}~\bibnamefont {Schulte}}, \bibinfo {author} {\bibfnamefont
  {D.}~\bibnamefont {Nitzschke}}, \bibinfo {author} {\bibfnamefont {M.~J.}\
  \bibnamefont {Borchert}}, \bibinfo {author} {\bibfnamefont {K.}~\bibnamefont
  {Hammerer}}, \bibinfo {author} {\bibfnamefont {S.}~\bibnamefont {Ulmer}}, \
  and\ \bibinfo {author} {\bibfnamefont {C.}~\bibnamefont {Ospelkaus}},\ }\href
  {\doibase 10.1088/1367-2630/ac136e} {\bibfield  {journal} {\bibinfo
  {journal} {arXiv preprint arXiv:2106.06252}\ } (\bibinfo {year} {2021}),\
  10.1088/1367-2630/ac136e},\ \Eprint {http://arxiv.org/abs/2106.06252}
  {arXiv:2106.06252} \BibitemShut {NoStop}%
\bibitem [{\citenamefont {Safronova}\ \emph {et~al.}(2018)\citenamefont
  {Safronova}, \citenamefont {Budker}, \citenamefont {DeMille}, \citenamefont
  {Kimball}, \citenamefont {Derevianko},\ and\ \citenamefont
  {Clark}}]{safronova2018}%
  \BibitemOpen
  \bibfield  {author} {\bibinfo {author} {\bibfnamefont {M.~S.}\ \bibnamefont
  {Safronova}}, \bibinfo {author} {\bibfnamefont {D.}~\bibnamefont {Budker}},
  \bibinfo {author} {\bibfnamefont {D.}~\bibnamefont {DeMille}}, \bibinfo
  {author} {\bibfnamefont {D.~F.~J.}\ \bibnamefont {Kimball}}, \bibinfo
  {author} {\bibfnamefont {A.}~\bibnamefont {Derevianko}}, \ and\ \bibinfo
  {author} {\bibfnamefont {C.~W.}\ \bibnamefont {Clark}},\ }\href {\doibase
  10.1103/RevModPhys.90.025008} {\bibfield  {journal} {\bibinfo  {journal}
  {Rev. Mod. Phys.}\ }\textbf {\bibinfo {volume} {90}},\ \bibinfo {pages}
  {025008} (\bibinfo {year} {2018})}\BibitemShut {NoStop}%
\bibitem [{\citenamefont {Schmidt}\ \emph {et~al.}(2005)\citenamefont
  {Schmidt}, \citenamefont {Rosenband}, \citenamefont {Langer}, \citenamefont
  {Itano}, \citenamefont {Bergquist},\ and\ \citenamefont
  {Wineland}}]{schmidt2005}%
  \BibitemOpen
  \bibfield  {author} {\bibinfo {author} {\bibfnamefont {P.~O.}\ \bibnamefont
  {Schmidt}}, \bibinfo {author} {\bibfnamefont {T.}~\bibnamefont {Rosenband}},
  \bibinfo {author} {\bibfnamefont {C.}~\bibnamefont {Langer}}, \bibinfo
  {author} {\bibfnamefont {W.~M.}\ \bibnamefont {Itano}}, \bibinfo {author}
  {\bibfnamefont {J.~C.}\ \bibnamefont {Bergquist}}, \ and\ \bibinfo {author}
  {\bibfnamefont {D.~J.}\ \bibnamefont {Wineland}},\ }\href {\doibase
  10.1126/science.1114375} {\bibfield  {journal} {\bibinfo  {journal}
  {Science}\ }\textbf {\bibinfo {volume} {309}},\ \bibinfo {pages} {749}
  (\bibinfo {year} {2005})}\BibitemShut {NoStop}%
\bibitem [{\citenamefont {Hume}\ \emph {et~al.}(2007)\citenamefont {Hume},
  \citenamefont {Rosenband},\ and\ \citenamefont {Wineland}}]{hume2007}%
  \BibitemOpen
  \bibfield  {author} {\bibinfo {author} {\bibfnamefont {D.~B.}\ \bibnamefont
  {Hume}}, \bibinfo {author} {\bibfnamefont {T.}~\bibnamefont {Rosenband}}, \
  and\ \bibinfo {author} {\bibfnamefont {D.~J.}\ \bibnamefont {Wineland}},\
  }\href {\doibase 10.1103/PhysRevLett.99.120502} {\bibfield  {journal}
  {\bibinfo  {journal} {Phys. Rev. Lett.}\ }\textbf {\bibinfo {volume} {99}},\
  \bibinfo {pages} {4} (\bibinfo {year} {2007})}\BibitemShut {NoStop}%
\bibitem [{\citenamefont {Beloy}\ \emph {et~al.}(2021)\citenamefont {Beloy},
  \citenamefont {Bodine}, \citenamefont {Bothwell}, \citenamefont {Brewer},
  \citenamefont {Bromley}, \citenamefont {Chen}, \citenamefont {Desch{\^e}nes},
  \citenamefont {Diddams}, \citenamefont {Fasano}, \citenamefont {Fortier},
  \citenamefont {Hassan}, \citenamefont {Hume}, \citenamefont {Kedar},
  \citenamefont {Kennedy}, \citenamefont {Khader}, \citenamefont {Koepke},
  \citenamefont {Leibrandt}, \citenamefont {Leopardi}, \citenamefont {Ludlow},
  \citenamefont {McGrew}, \citenamefont {Milner}, \citenamefont {Newbury},
  \citenamefont {Nicolodi}, \citenamefont {Oelker}, \citenamefont {Parker},
  \citenamefont {Robinson}, \citenamefont {Romisch}, \citenamefont
  {Sch{\"a}ffer}, \citenamefont {Sherman}, \citenamefont {Sinclair},
  \citenamefont {Sonderhouse}, \citenamefont {Swann}, \citenamefont {Yao},
  \citenamefont {Ye}, \citenamefont {Zhang},\ and\ \citenamefont {{Boulder
  Atomic Clock Optical Network (BACON) Collaboration*}}}]{beloy2021}%
  \BibitemOpen
  \bibfield  {author} {\bibinfo {author} {\bibfnamefont {K.}~\bibnamefont
  {Beloy}}, \bibinfo {author} {\bibfnamefont {M.~I.}\ \bibnamefont {Bodine}},
  \bibinfo {author} {\bibfnamefont {T.}~\bibnamefont {Bothwell}}, \bibinfo
  {author} {\bibfnamefont {S.~M.}\ \bibnamefont {Brewer}}, \bibinfo {author}
  {\bibfnamefont {S.~L.}\ \bibnamefont {Bromley}}, \bibinfo {author}
  {\bibfnamefont {J.-S.}\ \bibnamefont {Chen}}, \bibinfo {author}
  {\bibfnamefont {J.-D.}\ \bibnamefont {Desch{\^e}nes}}, \bibinfo {author}
  {\bibfnamefont {S.~A.}\ \bibnamefont {Diddams}}, \bibinfo {author}
  {\bibfnamefont {R.~J.}\ \bibnamefont {Fasano}}, \bibinfo {author}
  {\bibfnamefont {T.~M.}\ \bibnamefont {Fortier}}, \bibinfo {author}
  {\bibfnamefont {Y.~S.}\ \bibnamefont {Hassan}}, \bibinfo {author}
  {\bibfnamefont {D.~B.}\ \bibnamefont {Hume}}, \bibinfo {author}
  {\bibfnamefont {D.}~\bibnamefont {Kedar}}, \bibinfo {author} {\bibfnamefont
  {C.~J.}\ \bibnamefont {Kennedy}}, \bibinfo {author} {\bibfnamefont
  {I.}~\bibnamefont {Khader}}, \bibinfo {author} {\bibfnamefont
  {A.}~\bibnamefont {Koepke}}, \bibinfo {author} {\bibfnamefont {D.~R.}\
  \bibnamefont {Leibrandt}}, \bibinfo {author} {\bibfnamefont {H.}~\bibnamefont
  {Leopardi}}, \bibinfo {author} {\bibfnamefont {A.~D.}\ \bibnamefont
  {Ludlow}}, \bibinfo {author} {\bibfnamefont {W.~F.}\ \bibnamefont {McGrew}},
  \bibinfo {author} {\bibfnamefont {W.~R.}\ \bibnamefont {Milner}}, \bibinfo
  {author} {\bibfnamefont {N.~R.}\ \bibnamefont {Newbury}}, \bibinfo {author}
  {\bibfnamefont {D.}~\bibnamefont {Nicolodi}}, \bibinfo {author}
  {\bibfnamefont {E.}~\bibnamefont {Oelker}}, \bibinfo {author} {\bibfnamefont
  {T.~E.}\ \bibnamefont {Parker}}, \bibinfo {author} {\bibfnamefont {J.~M.}\
  \bibnamefont {Robinson}}, \bibinfo {author} {\bibfnamefont {S.}~\bibnamefont
  {Romisch}}, \bibinfo {author} {\bibfnamefont {S.~A.}\ \bibnamefont
  {Sch{\"a}ffer}}, \bibinfo {author} {\bibfnamefont {J.~A.}\ \bibnamefont
  {Sherman}}, \bibinfo {author} {\bibfnamefont {L.~C.}\ \bibnamefont
  {Sinclair}}, \bibinfo {author} {\bibfnamefont {L.}~\bibnamefont
  {Sonderhouse}}, \bibinfo {author} {\bibfnamefont {W.~C.}\ \bibnamefont
  {Swann}}, \bibinfo {author} {\bibfnamefont {J.}~\bibnamefont {Yao}}, \bibinfo
  {author} {\bibfnamefont {J.}~\bibnamefont {Ye}}, \bibinfo {author}
  {\bibfnamefont {X.}~\bibnamefont {Zhang}}, \ and\ \bibinfo {author}
  {\bibnamefont {{Boulder Atomic Clock Optical Network (BACON)
  Collaboration*}}},\ }\href {\doibase 10.1038/s41586-021-03253-4} {\bibfield
  {journal} {\bibinfo  {journal} {Nature}\ }\textbf {\bibinfo {volume} {591}},\
  \bibinfo {pages} {564} (\bibinfo {year} {2021})}\BibitemShut {NoStop}%
\bibitem [{\citenamefont {Hannig}\ \emph {et~al.}(2019)\citenamefont {Hannig},
  \citenamefont {Pelzer}, \citenamefont {Scharnhorst}, \citenamefont {Kramer},
  \citenamefont {Stepanova}, \citenamefont {Xu}, \citenamefont {Spethmann},
  \citenamefont {Leroux}, \citenamefont {Mehlst{\"a}ubler},\ and\ \citenamefont
  {Schmidt}}]{hannig2019}%
  \BibitemOpen
  \bibfield  {author} {\bibinfo {author} {\bibfnamefont {S.}~\bibnamefont
  {Hannig}}, \bibinfo {author} {\bibfnamefont {L.}~\bibnamefont {Pelzer}},
  \bibinfo {author} {\bibfnamefont {N.}~\bibnamefont {Scharnhorst}}, \bibinfo
  {author} {\bibfnamefont {J.}~\bibnamefont {Kramer}}, \bibinfo {author}
  {\bibfnamefont {M.}~\bibnamefont {Stepanova}}, \bibinfo {author}
  {\bibfnamefont {Z.~T.}\ \bibnamefont {Xu}}, \bibinfo {author} {\bibfnamefont
  {N.}~\bibnamefont {Spethmann}}, \bibinfo {author} {\bibfnamefont {I.~D.}\
  \bibnamefont {Leroux}}, \bibinfo {author} {\bibfnamefont {T.~E.}\
  \bibnamefont {Mehlst{\"a}ubler}}, \ and\ \bibinfo {author} {\bibfnamefont
  {P.~O.}\ \bibnamefont {Schmidt}},\ }\href {\doibase 10.1063/1.5090583}
  {\bibfield  {journal} {\bibinfo  {journal} {Rev. Sci. Instrum.}\ }\textbf
  {\bibinfo {volume} {90}},\ \bibinfo {pages} {053204} (\bibinfo {year}
  {2019})}\BibitemShut {NoStop}%
\bibitem [{\citenamefont {Cui}\ \emph {et~al.}(2020)\citenamefont {Cui},
  \citenamefont {Chao}, \citenamefont {Sun}, \citenamefont {Wang},
  \citenamefont {Zhang}, \citenamefont {Wei}, \citenamefont {Cao},
  \citenamefont {Shu},\ and\ \citenamefont {Huang}}]{cui2020}%
  \BibitemOpen
  \bibfield  {author} {\bibinfo {author} {\bibfnamefont {K.}~\bibnamefont
  {Cui}}, \bibinfo {author} {\bibfnamefont {S.}~\bibnamefont {Chao}}, \bibinfo
  {author} {\bibfnamefont {C.}~\bibnamefont {Sun}}, \bibinfo {author}
  {\bibfnamefont {S.}~\bibnamefont {Wang}}, \bibinfo {author} {\bibfnamefont
  {P.}~\bibnamefont {Zhang}}, \bibinfo {author} {\bibfnamefont
  {Y.}~\bibnamefont {Wei}}, \bibinfo {author} {\bibfnamefont {J.}~\bibnamefont
  {Cao}}, \bibinfo {author} {\bibfnamefont {H.}~\bibnamefont {Shu}}, \ and\
  \bibinfo {author} {\bibfnamefont {X.}~\bibnamefont {Huang}},\ }\href@noop {}
  {\bibfield  {journal} {\bibinfo  {journal} {arXiv: 2012.05496}\ } (\bibinfo
  {year} {2020})}\BibitemShut {NoStop}%
\bibitem [{\citenamefont {Brewer}\ \emph {et~al.}(2019)\citenamefont {Brewer},
  \citenamefont {Chen}, \citenamefont {Hankin}, \citenamefont {Clements},
  \citenamefont {Chou}, \citenamefont {Wineland}, \citenamefont {Hume},\ and\
  \citenamefont {Leibrandt}}]{brewer2019}%
  \BibitemOpen
  \bibfield  {author} {\bibinfo {author} {\bibfnamefont {S.~M.}\ \bibnamefont
  {Brewer}}, \bibinfo {author} {\bibfnamefont {J.-S.}\ \bibnamefont {Chen}},
  \bibinfo {author} {\bibfnamefont {A.~M.}\ \bibnamefont {Hankin}}, \bibinfo
  {author} {\bibfnamefont {E.~R.}\ \bibnamefont {Clements}}, \bibinfo {author}
  {\bibfnamefont {C.~W.}\ \bibnamefont {Chou}}, \bibinfo {author}
  {\bibfnamefont {D.~J.}\ \bibnamefont {Wineland}}, \bibinfo {author}
  {\bibfnamefont {D.~B.}\ \bibnamefont {Hume}}, \ and\ \bibinfo {author}
  {\bibfnamefont {D.~R.}\ \bibnamefont {Leibrandt}},\ }\href {\doibase
  10.1103/PhysRevLett.123.033201} {\bibfield  {journal} {\bibinfo  {journal}
  {Phys. Rev. Lett.}\ }\textbf {\bibinfo {volume} {123}},\ \bibinfo {pages}
  {033201} (\bibinfo {year} {2019})}\BibitemShut {NoStop}%
\bibitem [{\citenamefont {Chou}\ \emph {et~al.}(2011)\citenamefont {Chou},
  \citenamefont {Hume}, \citenamefont {Thorpe}, \citenamefont {Wineland},\ and\
  \citenamefont {Rosenband}}]{chou2011}%
  \BibitemOpen
  \bibfield  {author} {\bibinfo {author} {\bibfnamefont {C.~W.}\ \bibnamefont
  {Chou}}, \bibinfo {author} {\bibfnamefont {D.~B.}\ \bibnamefont {Hume}},
  \bibinfo {author} {\bibfnamefont {M.~J.}\ \bibnamefont {Thorpe}}, \bibinfo
  {author} {\bibfnamefont {D.~J.}\ \bibnamefont {Wineland}}, \ and\ \bibinfo
  {author} {\bibfnamefont {T.}~\bibnamefont {Rosenband}},\ }\href {\doibase
  10.1103/PhysRevLett.106.160801} {\bibfield  {journal} {\bibinfo  {journal}
  {Phys. Rev. Lett.}\ }\textbf {\bibinfo {volume} {106}},\ \bibinfo {pages} {4}
  (\bibinfo {year} {2011})}\BibitemShut {NoStop}%
\bibitem [{\citenamefont {Hume}\ \emph {et~al.}(2011)\citenamefont {Hume},
  \citenamefont {Chou}, \citenamefont {Leibrandt}, \citenamefont {Thorpe},
  \citenamefont {Wineland},\ and\ \citenamefont {Rosenband}}]{hume2011}%
  \BibitemOpen
  \bibfield  {author} {\bibinfo {author} {\bibfnamefont {D.~B.}\ \bibnamefont
  {Hume}}, \bibinfo {author} {\bibfnamefont {C.~W.}\ \bibnamefont {Chou}},
  \bibinfo {author} {\bibfnamefont {D.~R.}\ \bibnamefont {Leibrandt}}, \bibinfo
  {author} {\bibfnamefont {M.~J.}\ \bibnamefont {Thorpe}}, \bibinfo {author}
  {\bibfnamefont {D.~J.}\ \bibnamefont {Wineland}}, \ and\ \bibinfo {author}
  {\bibfnamefont {T.}~\bibnamefont {Rosenband}},\ }\href {\doibase
  10.1103/PhysRevLett.107.243902} {\bibfield  {journal} {\bibinfo  {journal}
  {Phys. Rev. Lett.}\ }\textbf {\bibinfo {volume} {107}},\ \bibinfo {pages} {5}
  (\bibinfo {year} {2011})}\BibitemShut {NoStop}%
\bibitem [{\citenamefont {Tan}\ \emph {et~al.}(2015)\citenamefont {Tan},
  \citenamefont {Gaebler}, \citenamefont {Lin}, \citenamefont {Wan},
  \citenamefont {Bowler}, \citenamefont {Leibfried},\ and\ \citenamefont
  {Wineland}}]{tan2015}%
  \BibitemOpen
  \bibfield  {author} {\bibinfo {author} {\bibfnamefont {T.~R.}\ \bibnamefont
  {Tan}}, \bibinfo {author} {\bibfnamefont {J.~P.}\ \bibnamefont {Gaebler}},
  \bibinfo {author} {\bibfnamefont {Y.}~\bibnamefont {Lin}}, \bibinfo {author}
  {\bibfnamefont {Y.}~\bibnamefont {Wan}}, \bibinfo {author} {\bibfnamefont
  {R.}~\bibnamefont {Bowler}}, \bibinfo {author} {\bibfnamefont
  {D.}~\bibnamefont {Leibfried}}, \ and\ \bibinfo {author} {\bibfnamefont
  {D.~J.}\ \bibnamefont {Wineland}},\ }\href@noop {} {\bibfield  {journal}
  {\bibinfo  {journal} {Nature}\ }\textbf {\bibinfo {volume} {528}},\ \bibinfo
  {pages} {380} (\bibinfo {year} {2015})}\BibitemShut {NoStop}%
\bibitem [{\citenamefont {Kienzler}\ \emph {et~al.}(2020)\citenamefont
  {Kienzler}, \citenamefont {Wan}, \citenamefont {Erickson}, \citenamefont
  {Wu}, \citenamefont {Wilson}, \citenamefont {Wineland},\ and\ \citenamefont
  {Leibfried}}]{kienzler2020}%
  \BibitemOpen
  \bibfield  {author} {\bibinfo {author} {\bibfnamefont {D.}~\bibnamefont
  {Kienzler}}, \bibinfo {author} {\bibfnamefont {Y.}~\bibnamefont {Wan}},
  \bibinfo {author} {\bibfnamefont {S.~D.}\ \bibnamefont {Erickson}}, \bibinfo
  {author} {\bibfnamefont {J.~J.}\ \bibnamefont {Wu}}, \bibinfo {author}
  {\bibfnamefont {A.~C.}\ \bibnamefont {Wilson}}, \bibinfo {author}
  {\bibfnamefont {D.~J.}\ \bibnamefont {Wineland}}, \ and\ \bibinfo {author}
  {\bibfnamefont {D.}~\bibnamefont {Leibfried}},\ }\href {\doibase
  10.1103/PhysRevX.10.021012} {\bibfield  {journal} {\bibinfo  {journal} {Phys.
  Rev. X}\ }\textbf {\bibinfo {volume} {10}},\ \bibinfo {pages} {021012}
  (\bibinfo {year} {2020})}\BibitemShut {NoStop}%
\bibitem [{\citenamefont {Schulte}\ \emph {et~al.}(2016)\citenamefont
  {Schulte}, \citenamefont {L\"{o}rch}, \citenamefont {Leroux}, \citenamefont
  {Schmidt},\ and\ \citenamefont {Hammerer}}]{schulte2016}%
  \BibitemOpen
  \bibfield  {author} {\bibinfo {author} {\bibfnamefont {M.}~\bibnamefont
  {Schulte}}, \bibinfo {author} {\bibfnamefont {N.}~\bibnamefont {L\"{o}rch}},
  \bibinfo {author} {\bibfnamefont {I.~D.}\ \bibnamefont {Leroux}}, \bibinfo
  {author} {\bibfnamefont {P.~O.}\ \bibnamefont {Schmidt}}, \ and\ \bibinfo
  {author} {\bibfnamefont {K.}~\bibnamefont {Hammerer}},\ }\href {\doibase
  10.1103/PhysRevLett.116.013002} {\bibfield  {journal} {\bibinfo  {journal}
  {Phys. Rev. Lett.}\ }\textbf {\bibinfo {volume} {116}} (\bibinfo {year}
  {2016}),\ 10.1103/PhysRevLett.116.013002}\BibitemShut {NoStop}%
\bibitem [{\citenamefont {Chen}\ \emph {et~al.}(2020)\citenamefont {Chen},
  \citenamefont {Wright}, \citenamefont {Pisenti}, \citenamefont {Murphy},
  \citenamefont {Beck}, \citenamefont {Landsman}, \citenamefont {Amini},\ and\
  \citenamefont {Nam}}]{chen2020}%
  \BibitemOpen
  \bibfield  {author} {\bibinfo {author} {\bibfnamefont {J.-S.}\ \bibnamefont
  {Chen}}, \bibinfo {author} {\bibfnamefont {K.}~\bibnamefont {Wright}},
  \bibinfo {author} {\bibfnamefont {N.~C.}\ \bibnamefont {Pisenti}}, \bibinfo
  {author} {\bibfnamefont {D.}~\bibnamefont {Murphy}}, \bibinfo {author}
  {\bibfnamefont {K.~M.}\ \bibnamefont {Beck}}, \bibinfo {author}
  {\bibfnamefont {K.}~\bibnamefont {Landsman}}, \bibinfo {author}
  {\bibfnamefont {J.~M.}\ \bibnamefont {Amini}}, \ and\ \bibinfo {author}
  {\bibfnamefont {Y.}~\bibnamefont {Nam}},\ }\href {\doibase
  10.1103/PhysRevA.102.043110} {\bibfield  {journal} {\bibinfo  {journal}
  {Phys. Rev. A}\ }\textbf {\bibinfo {volume} {102}},\ \bibinfo {pages}
  {043110} (\bibinfo {year} {2020})}\BibitemShut {NoStop}%
\bibitem [{\citenamefont {Feng}\ \emph {et~al.}(2020)\citenamefont {Feng},
  \citenamefont {Tan}, \citenamefont {De}, \citenamefont {Menon}, \citenamefont
  {Chu}, \citenamefont {Pagano},\ and\ \citenamefont {Monroe}}]{feng2020}%
  \BibitemOpen
  \bibfield  {author} {\bibinfo {author} {\bibfnamefont {L.}~\bibnamefont
  {Feng}}, \bibinfo {author} {\bibfnamefont {W.~L.}\ \bibnamefont {Tan}},
  \bibinfo {author} {\bibfnamefont {A.}~\bibnamefont {De}}, \bibinfo {author}
  {\bibfnamefont {A.}~\bibnamefont {Menon}}, \bibinfo {author} {\bibfnamefont
  {A.}~\bibnamefont {Chu}}, \bibinfo {author} {\bibfnamefont {G.}~\bibnamefont
  {Pagano}}, \ and\ \bibinfo {author} {\bibfnamefont {C.}~\bibnamefont
  {Monroe}},\ }\href {\doibase 10.1103/PhysRevLett.125.053001} {\bibfield
  {journal} {\bibinfo  {journal} {Phys. Rev. Lett.}\ }\textbf {\bibinfo
  {volume} {125}},\ \bibinfo {pages} {053001} (\bibinfo {year}
  {2020})}\BibitemShut {NoStop}%
\bibitem [{\citenamefont {Monroe}\ \emph {et~al.}(1996)\citenamefont {Monroe},
  \citenamefont {Meekhof}, \citenamefont {King},\ and\ \citenamefont
  {Wineland}}]{monroe1996}%
  \BibitemOpen
  \bibfield  {author} {\bibinfo {author} {\bibfnamefont {C.}~\bibnamefont
  {Monroe}}, \bibinfo {author} {\bibfnamefont {D.~M.}\ \bibnamefont {Meekhof}},
  \bibinfo {author} {\bibfnamefont {B.~E.}\ \bibnamefont {King}}, \ and\
  \bibinfo {author} {\bibfnamefont {D.~J.}\ \bibnamefont {Wineland}},\ }\href
  {\doibase 10.1126/science.272.5265.1131} {\bibfield  {journal} {\bibinfo
  {journal} {Science}\ }\textbf {\bibinfo {volume} {272}},\ \bibinfo {pages}
  {1131} (\bibinfo {year} {1996})}\BibitemShut {NoStop}%
\bibitem [{\citenamefont {Milne}\ \emph {et~al.}(2021)\citenamefont {Milne},
  \citenamefont {Hempel}, \citenamefont {Li}, \citenamefont {Edmunds},
  \citenamefont {Slatyer}, \citenamefont {Ball}, \citenamefont {Hush},\ and\
  \citenamefont {Biercuk}}]{milne2021}%
  \BibitemOpen
  \bibfield  {author} {\bibinfo {author} {\bibfnamefont {A.~R.}\ \bibnamefont
  {Milne}}, \bibinfo {author} {\bibfnamefont {C.}~\bibnamefont {Hempel}},
  \bibinfo {author} {\bibfnamefont {L.}~\bibnamefont {Li}}, \bibinfo {author}
  {\bibfnamefont {C.~L.}\ \bibnamefont {Edmunds}}, \bibinfo {author}
  {\bibfnamefont {H.~J.}\ \bibnamefont {Slatyer}}, \bibinfo {author}
  {\bibfnamefont {H.}~\bibnamefont {Ball}}, \bibinfo {author} {\bibfnamefont
  {M.~R.}\ \bibnamefont {Hush}}, \ and\ \bibinfo {author} {\bibfnamefont
  {M.~J.}\ \bibnamefont {Biercuk}},\ }\href {\doibase
  10.1103/PhysRevLett.126.250506} {\bibfield  {journal} {\bibinfo  {journal}
  {Phys. Rev. Lett.}\ }\textbf {\bibinfo {volume} {126}},\ \bibinfo {pages}
  {250506} (\bibinfo {year} {2021})}\BibitemShut {NoStop}%
\bibitem [{\citenamefont {Haljan}\ \emph {et~al.}(2005)\citenamefont {Haljan},
  \citenamefont {Brickman}, \citenamefont {Deslauriers}, \citenamefont {Lee},\
  and\ \citenamefont {Monroe}}]{haljan2005}%
  \BibitemOpen
  \bibfield  {author} {\bibinfo {author} {\bibfnamefont {P.~C.}\ \bibnamefont
  {Haljan}}, \bibinfo {author} {\bibfnamefont {K.-A.}\ \bibnamefont
  {Brickman}}, \bibinfo {author} {\bibfnamefont {L.}~\bibnamefont
  {Deslauriers}}, \bibinfo {author} {\bibfnamefont {P.~J.}\ \bibnamefont
  {Lee}}, \ and\ \bibinfo {author} {\bibfnamefont {C.}~\bibnamefont {Monroe}},\
  }\href {\doibase 10.1103/PhysRevLett.94.153602} {\bibfield  {journal}
  {\bibinfo  {journal} {Phys. Rev. Lett.}\ }\textbf {\bibinfo {volume} {94}},\
  \bibinfo {pages} {153602} (\bibinfo {year} {2005})}\BibitemShut {NoStop}%
\bibitem [{\citenamefont {Hempel}\ \emph {et~al.}(2013)\citenamefont {Hempel},
  \citenamefont {Lanyon}, \citenamefont {Jurcevic}, \citenamefont {Gerritsma},
  \citenamefont {Blatt},\ and\ \citenamefont {Roos}}]{hempel2013}%
  \BibitemOpen
  \bibfield  {author} {\bibinfo {author} {\bibfnamefont {C.}~\bibnamefont
  {Hempel}}, \bibinfo {author} {\bibfnamefont {B.~P.}\ \bibnamefont {Lanyon}},
  \bibinfo {author} {\bibfnamefont {P.}~\bibnamefont {Jurcevic}}, \bibinfo
  {author} {\bibfnamefont {R.}~\bibnamefont {Gerritsma}}, \bibinfo {author}
  {\bibfnamefont {R.}~\bibnamefont {Blatt}}, \ and\ \bibinfo {author}
  {\bibfnamefont {C.~F.}\ \bibnamefont {Roos}},\ }\href {\doibase
  10.1038/nphoton.2013.172} {\bibfield  {journal} {\bibinfo  {journal} {Nat.
  Photonics}\ }\textbf {\bibinfo {volume} {7}},\ \bibinfo {pages} {630}
  (\bibinfo {year} {2013})}\BibitemShut {NoStop}%
\bibitem [{\citenamefont {M{\o}lmer}\ and\ \citenamefont
  {S{\o}rensen}(1999)}]{molmer1999}%
  \BibitemOpen
  \bibfield  {author} {\bibinfo {author} {\bibfnamefont {K.}~\bibnamefont
  {M{\o}lmer}}\ and\ \bibinfo {author} {\bibfnamefont {A.}~\bibnamefont
  {S{\o}rensen}},\ }\href {\doibase 10.1103/PhysRevLett.82.1835} {\bibfield
  {journal} {\bibinfo  {journal} {Phys. Rev. Lett.}\ }\textbf {\bibinfo
  {volume} {82}},\ \bibinfo {pages} {1835} (\bibinfo {year}
  {1999})}\BibitemShut {NoStop}%
\bibitem [{\citenamefont {Wineland}\ \emph {et~al.}(1998)\citenamefont
  {Wineland}, \citenamefont {Monroe}, \citenamefont {Itano}, \citenamefont
  {Leibfried}, \citenamefont {King},\ and\ \citenamefont
  {Meekhof}}]{wineland1998}%
  \BibitemOpen
  \bibfield  {author} {\bibinfo {author} {\bibfnamefont {D.~J.}\ \bibnamefont
  {Wineland}}, \bibinfo {author} {\bibfnamefont {C.}~\bibnamefont {Monroe}},
  \bibinfo {author} {\bibfnamefont {W.~M.}\ \bibnamefont {Itano}}, \bibinfo
  {author} {\bibfnamefont {D.}~\bibnamefont {Leibfried}}, \bibinfo {author}
  {\bibfnamefont {B.~E.}\ \bibnamefont {King}}, \ and\ \bibinfo {author}
  {\bibfnamefont {D.~M.}\ \bibnamefont {Meekhof}},\ }\href {\doibase
  10.6028/jres.103.019} {\bibfield  {journal} {\bibinfo  {journal} {J. Res.
  Natl. Inst. Stan.}\ }\textbf {\bibinfo {volume} {103}},\ \bibinfo {pages}
  {259} (\bibinfo {year} {1998})}\BibitemShut {NoStop}%
\bibitem [{Note100()}]{Note100}%
  \BibitemOpen
  \bibinfo {note} {See supplemental material}\BibitemShut {NoStop}%
\bibitem [{\citenamefont {Chen}\ \emph {et~al.}(2017)\citenamefont {Chen},
  \citenamefont {Brewer}, \citenamefont {Chou}, \citenamefont {Wineland},
  \citenamefont {Leibrandt},\ and\ \citenamefont {Hume}}]{chen2017}%
  \BibitemOpen
  \bibfield  {author} {\bibinfo {author} {\bibfnamefont {J.-S.}\ \bibnamefont
  {Chen}}, \bibinfo {author} {\bibfnamefont {S.~M.}\ \bibnamefont {Brewer}},
  \bibinfo {author} {\bibfnamefont {C.}~\bibnamefont {Chou}}, \bibinfo {author}
  {\bibfnamefont {D.}~\bibnamefont {Wineland}}, \bibinfo {author}
  {\bibfnamefont {D.}~\bibnamefont {Leibrandt}}, \ and\ \bibinfo {author}
  {\bibfnamefont {D.}~\bibnamefont {Hume}},\ }\href@noop {} {\bibfield
  {journal} {\bibinfo  {journal} {Phys. Rev. Lett.}\ }\textbf {\bibinfo
  {volume} {118}},\ \bibinfo {pages} {053002} (\bibinfo {year}
  {2017})}\BibitemShut {NoStop}%
\bibitem [{\citenamefont {Rosenband}\ \emph {et~al.}(2007)\citenamefont
  {Rosenband}, \citenamefont {Schmidt}, \citenamefont {Hume}, \citenamefont
  {Itano}, \citenamefont {Fortier}, \citenamefont {Stalnaker}, \citenamefont
  {Kim}, \citenamefont {Diddams}, \citenamefont {Koelemeij}, \citenamefont
  {Bergquist},\ and\ \citenamefont {Wineland}}]{rosenband2007}%
  \BibitemOpen
  \bibfield  {author} {\bibinfo {author} {\bibfnamefont {T.}~\bibnamefont
  {Rosenband}}, \bibinfo {author} {\bibfnamefont {P.~O.}\ \bibnamefont
  {Schmidt}}, \bibinfo {author} {\bibfnamefont {D.~B.}\ \bibnamefont {Hume}},
  \bibinfo {author} {\bibfnamefont {W.~M.}\ \bibnamefont {Itano}}, \bibinfo
  {author} {\bibfnamefont {T.~M.}\ \bibnamefont {Fortier}}, \bibinfo {author}
  {\bibfnamefont {J.~E.}\ \bibnamefont {Stalnaker}}, \bibinfo {author}
  {\bibfnamefont {K.}~\bibnamefont {Kim}}, \bibinfo {author} {\bibfnamefont
  {S.~A.}\ \bibnamefont {Diddams}}, \bibinfo {author} {\bibfnamefont
  {J.~C.~J.}\ \bibnamefont {Koelemeij}}, \bibinfo {author} {\bibfnamefont
  {J.~C.}\ \bibnamefont {Bergquist}}, \ and\ \bibinfo {author} {\bibfnamefont
  {D.~J.}\ \bibnamefont {Wineland}},\ }\href {\doibase
  10.1103/PhysRevLett.98.220801} {\bibfield  {journal} {\bibinfo  {journal}
  {Phys. Rev. Lett.}\ }\textbf {\bibinfo {volume} {98}},\ \bibinfo {pages}
  {220801} (\bibinfo {year} {2007})}\BibitemShut {NoStop}%
\bibitem [{\citenamefont {Erickson}\ \emph {et~al.}(2021)\citenamefont
  {Erickson}, \citenamefont {Wu}, \citenamefont {Hou}, \citenamefont {Cole},
  \citenamefont {Geller}, \citenamefont {Kwiatkowski}, \citenamefont {Glancy},
  \citenamefont {Knill}, \citenamefont {Slichter}, \citenamefont {Wilson} \emph
  {et~al.}}]{erickson2021}%
  \BibitemOpen
  \bibfield  {author} {\bibinfo {author} {\bibfnamefont {S.~D.}\ \bibnamefont
  {Erickson}}, \bibinfo {author} {\bibfnamefont {J.~J.}\ \bibnamefont {Wu}},
  \bibinfo {author} {\bibfnamefont {P.-Y.}\ \bibnamefont {Hou}}, \bibinfo
  {author} {\bibfnamefont {D.~C.}\ \bibnamefont {Cole}}, \bibinfo {author}
  {\bibfnamefont {S.}~\bibnamefont {Geller}}, \bibinfo {author} {\bibfnamefont
  {A.}~\bibnamefont {Kwiatkowski}}, \bibinfo {author} {\bibfnamefont
  {S.}~\bibnamefont {Glancy}}, \bibinfo {author} {\bibfnamefont
  {E.}~\bibnamefont {Knill}}, \bibinfo {author} {\bibfnamefont {D.~H.}\
  \bibnamefont {Slichter}}, \bibinfo {author} {\bibfnamefont {A.~C.}\
  \bibnamefont {Wilson}},  \emph {et~al.},\ }\href@noop {} {\bibfield
  {journal} {\bibinfo  {journal} {arXiv preprint arXiv:2112.06341}\ } (\bibinfo
  {year} {2021})}\BibitemShut {NoStop}%
\bibitem [{\citenamefont {Bollinger}\ \emph {et~al.}(1996)\citenamefont
  {Bollinger}, \citenamefont {Itano}, \citenamefont {Wineland},\ and\
  \citenamefont {Heinzen}}]{bollinger1996}%
  \BibitemOpen
  \bibfield  {author} {\bibinfo {author} {\bibfnamefont {J.~J.~.}\ \bibnamefont
  {Bollinger}}, \bibinfo {author} {\bibfnamefont {W.~M.}\ \bibnamefont
  {Itano}}, \bibinfo {author} {\bibfnamefont {D.~J.}\ \bibnamefont {Wineland}},
  \ and\ \bibinfo {author} {\bibfnamefont {D.~J.}\ \bibnamefont {Heinzen}},\
  }\href {\doibase 10.1103/PhysRevA.54.R4649} {\bibfield  {journal} {\bibinfo
  {journal} {Phys. Rev. A}\ }\textbf {\bibinfo {volume} {54}},\ \bibinfo
  {pages} {R4649} (\bibinfo {year} {1996})}\BibitemShut {NoStop}%
\bibitem [{\citenamefont {Gilmore}\ \emph {et~al.}(2017)\citenamefont
  {Gilmore}, \citenamefont {Bohnet}, \citenamefont {Sawyer}, \citenamefont
  {Britton},\ and\ \citenamefont {Bollinger}}]{gilmore2017}%
  \BibitemOpen
  \bibfield  {author} {\bibinfo {author} {\bibfnamefont {K.~A.}\ \bibnamefont
  {Gilmore}}, \bibinfo {author} {\bibfnamefont {J.~G.}\ \bibnamefont {Bohnet}},
  \bibinfo {author} {\bibfnamefont {B.~C.}\ \bibnamefont {Sawyer}}, \bibinfo
  {author} {\bibfnamefont {J.~W.}\ \bibnamefont {Britton}}, \ and\ \bibinfo
  {author} {\bibfnamefont {J.~J.}\ \bibnamefont {Bollinger}},\ }\href {\doibase
  10.1103/PhysRevLett.118.263602} {\bibfield  {journal} {\bibinfo  {journal}
  {Phys. Rev. Lett.}\ }\textbf {\bibinfo {volume} {118}},\ \bibinfo {pages}
  {263602} (\bibinfo {year} {2017})}\BibitemShut {NoStop}%
\bibitem [{\citenamefont {Affolter}\ \emph {et~al.}(2020)\citenamefont
  {Affolter}, \citenamefont {Gilmore}, \citenamefont {Jordan},\ and\
  \citenamefont {Bollinger}}]{affolter2020}%
  \BibitemOpen
  \bibfield  {author} {\bibinfo {author} {\bibfnamefont {M.}~\bibnamefont
  {Affolter}}, \bibinfo {author} {\bibfnamefont {K.~A.}\ \bibnamefont
  {Gilmore}}, \bibinfo {author} {\bibfnamefont {J.~E.}\ \bibnamefont {Jordan}},
  \ and\ \bibinfo {author} {\bibfnamefont {J.~J.}\ \bibnamefont {Bollinger}},\
  }\href {\doibase 10.1103/PhysRevA.102.052609} {\bibfield  {journal} {\bibinfo
   {journal} {Phys. Rev. A}\ }\textbf {\bibinfo {volume} {102}},\ \bibinfo
  {pages} {052609} (\bibinfo {year} {2020})}\BibitemShut {NoStop}%
\bibitem [{\citenamefont {Arecchi}\ \emph {et~al.}(1972)\citenamefont
  {Arecchi}, \citenamefont {Courtens}, \citenamefont {Gilmore},\ and\
  \citenamefont {Thomas}}]{arecchi1972}%
  \BibitemOpen
  \bibfield  {author} {\bibinfo {author} {\bibfnamefont {F.~T.}\ \bibnamefont
  {Arecchi}}, \bibinfo {author} {\bibfnamefont {E.}~\bibnamefont {Courtens}},
  \bibinfo {author} {\bibfnamefont {R.}~\bibnamefont {Gilmore}}, \ and\
  \bibinfo {author} {\bibfnamefont {H.}~\bibnamefont {Thomas}},\ }\href
  {\doibase 10.1103/PhysRevA.6.2211} {\bibfield  {journal} {\bibinfo  {journal}
  {Phys. Rev. A}\ }\textbf {\bibinfo {volume} {6}},\ \bibinfo {pages} {2211}
  (\bibinfo {year} {1972})}\BibitemShut {NoStop}%
\end{thebibliography}%


\begin{thebibliography}{3}%
\makeatletter
\providecommand \@ifxundefined [1]{%
 \@ifx{#1\undefined}
}%
\providecommand \@ifnum [1]{%
 \ifnum #1\expandafter \@firstoftwo
 \else \expandafter \@secondoftwo
 \fi
}%
\providecommand \@ifx [1]{%
 \ifx #1\expandafter \@firstoftwo
 \else \expandafter \@secondoftwo
 \fi
}%
\providecommand \natexlab [1]{#1}%
\providecommand \enquote  [1]{``#1''}%
\providecommand \bibnamefont  [1]{#1}%
\providecommand \bibfnamefont [1]{#1}%
\providecommand \citenamefont [1]{#1}%
\providecommand \href@noop [0]{\@secondoftwo}%
\providecommand \href [0]{\begingroup \@sanitize@url \@href}%
\providecommand \@href[1]{\@@startlink{#1}\@@href}%
\providecommand \@@href[1]{\endgroup#1\@@endlink}%
\providecommand \@sanitize@url [0]{\catcode `\\12\catcode `\$12\catcode
  `\&12\catcode `\#12\catcode `\^12\catcode `\_12\catcode `\%12\relax}%
\providecommand \@@startlink[1]{}%
\providecommand \@@endlink[0]{}%
\providecommand \url  [0]{\begingroup\@sanitize@url \@url }%
\providecommand \@url [1]{\endgroup\@href {#1}{\urlprefix }}%
\providecommand \urlprefix  [0]{URL }%
\providecommand \Eprint [0]{\href }%
\providecommand \doibase [0]{http://dx.doi.org/}%
\providecommand \selectlanguage [0]{\@gobble}%
\providecommand \bibinfo  [0]{\@secondoftwo}%
\providecommand \bibfield  [0]{\@secondoftwo}%
\providecommand \translation [1]{[#1]}%
\providecommand \BibitemOpen [0]{}%
\providecommand \bibitemStop [0]{}%
\providecommand \bibitemNoStop [0]{.\EOS\space}%
\providecommand \EOS [0]{\spacefactor3000\relax}%
\providecommand \BibitemShut  [1]{\csname bibitem#1\endcsname}%
\let\auto@bib@innerbib\@empty
\bibitem [{\citenamefont {Arecchi}\ \emph {et~al.}(1972)\citenamefont
  {Arecchi}, \citenamefont {Courtens}, \citenamefont {Gilmore},\ and\
  \citenamefont {Thomas}}]{arecchi1972}%
  \BibitemOpen
  \bibfield  {author} {\bibinfo {author} {\bibfnamefont {F.~T.}\ \bibnamefont
  {Arecchi}}, \bibinfo {author} {\bibfnamefont {E.}~\bibnamefont {Courtens}},
  \bibinfo {author} {\bibfnamefont {R.}~\bibnamefont {Gilmore}}, \ and\
  \bibinfo {author} {\bibfnamefont {H.}~\bibnamefont {Thomas}},\ }\href
  {\doibase 10.1103/PhysRevA.6.2211} {\bibfield  {journal} {\bibinfo  {journal}
  {Phys. Rev. A}\ }\textbf {\bibinfo {volume} {6}},\ \bibinfo {pages} {2211}
  (\bibinfo {year} {1972})}\BibitemShut {NoStop}%
\bibitem [{\citenamefont {Wineland}\ \emph {et~al.}(1998)\citenamefont
  {Wineland}, \citenamefont {Monroe}, \citenamefont {Itano}, \citenamefont
  {Leibfried}, \citenamefont {King},\ and\ \citenamefont
  {Meekhof}}]{wineland1998}%
  \BibitemOpen
  \bibfield  {author} {\bibinfo {author} {\bibfnamefont {D.~J.}\ \bibnamefont
  {Wineland}}, \bibinfo {author} {\bibfnamefont {C.}~\bibnamefont {Monroe}},
  \bibinfo {author} {\bibfnamefont {W.~M.}\ \bibnamefont {Itano}}, \bibinfo
  {author} {\bibfnamefont {D.}~\bibnamefont {Leibfried}}, \bibinfo {author}
  {\bibfnamefont {B.~E.}\ \bibnamefont {King}}, \ and\ \bibinfo {author}
  {\bibfnamefont {D.~M.}\ \bibnamefont {Meekhof}},\ }\href {\doibase
  10.6028/jres.103.019} {\bibfield  {journal} {\bibinfo  {journal} {J. Res.
  Natl. Inst. Stan.}\ }\textbf {\bibinfo {volume} {103}},\ \bibinfo {pages}
  {259} (\bibinfo {year} {1998})}\BibitemShut {NoStop}%
\bibitem [{\citenamefont {Haljan}\ \emph {et~al.}(2005)\citenamefont {Haljan},
  \citenamefont {Brickman}, \citenamefont {Deslauriers}, \citenamefont {Lee},\
  and\ \citenamefont {Monroe}}]{haljan2005}%
  \BibitemOpen
  \bibfield  {author} {\bibinfo {author} {\bibfnamefont {P.~C.}\ \bibnamefont
  {Haljan}}, \bibinfo {author} {\bibfnamefont {K.-A.}\ \bibnamefont
  {Brickman}}, \bibinfo {author} {\bibfnamefont {L.}~\bibnamefont
  {Deslauriers}}, \bibinfo {author} {\bibfnamefont {P.~J.}\ \bibnamefont
  {Lee}}, \ and\ \bibinfo {author} {\bibfnamefont {C.}~\bibnamefont {Monroe}},\
  }\href {\doibase 10.1103/PhysRevLett.94.153602} {\bibfield  {journal}
  {\bibinfo  {journal} {Phys. Rev. Lett.}\ }\textbf {\bibinfo {volume} {94}},\
  \bibinfo {pages} {153602} (\bibinfo {year} {2005})}\BibitemShut {NoStop}%
\end{thebibliography}%

\end{document}


\raggedbottom

\title{\textit{Supplemental Material for}\\
    Scalable quantum logic spectroscopy}

\author{Kaifeng Cui}
\email{cuikaifeng@apm.ac.cn}
\affiliation{\NIST} 
\affiliation{\ANL}
\affiliation{\APM} %

\author{Jose Valencia}
\affiliation{\NIST}
\affiliation{\CU}

\author{Kevin T. Boyce}
\affiliation{\NIST}
\affiliation{\CU}

\author{Ethan R. Clements}
\affiliation{\NIST}
\affiliation{\CU}

\author{David R. Leibrandt}
\affiliation{\NIST}
\affiliation{\CU}

\author{David B. Hume}
\email{david.hume@nist.gov}
\affiliation{\NIST}

\date{\today}

\maketitle

\section{Detection signal of the Schr\"{o}dinger cat interferometer}

We start by considering a single logic ion (LI) coupled to a single spectroscopy ion (SI) through a shared mode of motion at frequency $\omega_M$.  
For the $\dl$ and $\ul$ states of the logic ion, we consider equal and opposite forces with amplitudes $\fdl = - \ful \equiv \fl$.
Similarly, for the spectroscopy system, with states $\ds$ and $\us$ we define $\fds = -\fus \equiv \fs$.  
For both LI and SI, $\ketda$ and $\ketua$ represent the eigenstates of $\hat{\sigma}_i$ Paul operator.

At the beginning, the logic ion is prepared in an equal superposition $\ket{\psi}_L = \left(\dl + \ul\right)/\sqrt{2}$ while the spectroscopy ion is assumed to be in state $\ds$. 
Application of the state-dependent force $F(t) = \pm\fl\cos\left(\omega_M t \right)$ for time $t_L$ results in state-dependent displacements (LI-SDDs) $\hat{\mathcal{D}}(\alpha)$ and $\hat{\mathcal{D}}(-\alpha)$ for the $\ul$ and $\dl$ states, respectively. 
The displacement amplitude is given by $\alpha_0 =|\alpha(t_L)| = F_L z_L t_L /2\hbar$, where $z_L$ is the extent of the logic ion wavefunction in the ground state.  
This results in a Schr\"{o}dinger cat state,
\begin{equation}
\ket{\psi} = \frac{1}{\sqrt{2}}\left(\dl\ket{+\alpha}_M+\ul\ket{-\alpha}_M\right)\ds,
\end{equation}
where the logic ion state is entangled with the motion.

A subsequent state-dependent force $F(t) = \pm\fs\cos\left(\omega_M t+\phi_M\right)$ applied to the spectroscopy ion for time $t_S$ results in another state-dependent displacement (SI-SDD) $D(\beta)$ at an angle $\phi_M$ relative to the initial wave-packet separation.
Here, equivalently to the logic ion case, we have $\beta_0 = |\beta(t_S)| = F_S z_S t_S/2\hbar$, where $z_S$ is the extent of the spectroscopy ion wave function in the ground state. 

The interferometer is finally closed by reversing the original LI-SDD (marked as LI-SDD$^{-1}$) via application of $F(t) = \mp\fl\cos\left(\omega_M t\right)$.
Note that a phase shift of $\pi$ is introduced.
Using $\hat{\mathcal{D}}(\beta)\hat{\mathcal{D}}(\alpha) = \hat{\mathcal{D}}(\beta+\alpha)e^{i|\alpha| |\beta|\rm{sin}\phi_M}$, we can write the state after all three displacements:
\begin{equation}
    \ket{\psi} = \frac{1}{\sqrt{2}}\left(
         e^{-2i\alpha_0 \beta_0\rm{sin}\phi_M}\dl
        +e^{ 2i\alpha_0 \beta_0\rm{sin}\phi_M}\ul
    \right) \ds \ket{\beta}_M,
\end{equation}
where we can see a relative phase $2\theta=4 \alpha_0 \beta_0 \rm{sin}\phi_M$
has been accumulated in this interferometer between the $\dl$ and $\ul$ state.

We can map the phase to populations by rotating our spin basis from $\hat{\sigma}_{L,i}$
to another orthogonal basis $\hat{\sigma}_{L,j}$ using 
${\ketda_{L,j} \rightarrow(\dl+\ul)/\sqrt{2}}$ and 
${\ketua_{L,j} \rightarrow(\dl-\ul)/\sqrt{2}}$.  
The resulting logic ion state is
\begin{equation}
\ket{\psi}_L = \cos(\theta)\ketda_{L,j} + i \sin(\theta)\ketua_{L,j}
\end{equation}
where we can observe state populations,
\begin{equation}
\begin{split}\label{eq:PopDet}
P_{\downarrow, L} = \rm{cos}^2(\theta) = (1+\rm{cos}(2\theta))/2,\\
P_{\uparrow, L} = \rm{sin}^2(\theta)= (1-\rm{cos}(2\theta))/2.\\
\end{split}
\end{equation}

This sinusoidal response can be used as a means of state detection for the spectroscopy ion system.  
Here, we suppose that a separate spectroscopy pulse, before the detection sequence, probes an auxiliary resonance $\ds\leftrightarrow\as$. 
To maximize signal-to-noise in detecting the transition $\ds\rightarrow\as$, we can choose parameters $\theta = \pi/2$
such that we get maximum transition probability $P_{\uparrow, L} = 1$ when the spectroscopy ion remains in state $\ds$.

\section{Scaling to multiple logic ions}

When applying spin-dependent forces on multiple ions simultaneously, we rely on several simplifying conditions that can be met in our experiment.
First, 
    we assume equal force amplitudes on all logic ions in the ion chain.  
    In our case that means that the laser beams used to produce an SDD have nearly equal intensity across the array.  
Second, 
    we require that all the ions have nearly equal mode amplitudes for the mode of motion addressed.  
These two conditions together ensure that the characteristic displacement amplitude, $|\alpha|$, for each logic ion is equal.  
Third, 
    we assume that the phase of the force is equal for all the ions.  Depending on the method chosen to produce the state-dependent forces, this can be satisfied experimentally to a greater or lesser degree.

With these conditions, the displacement amplitude for a particular logic ion state during the opening and closing pulses of the interferometer is given by $\alpha = (N_{L,\uparrow} - N_{L,\downarrow}) \alpha_0$.  
Here $N_{L,\uparrow}$ and $N_{L,\downarrow}$ count the number of logic ions in the states $\dl$ and $\ul$ respectively. 
With strictly resonant interactions in terms of the motional frequency and qubit frequency, we show that these conditions result in each logic ion acting as an independent sensor of the displacement applied to the spectroscopy ions.

We expand the state of the logic ion system in terms of the collective angular momentum states $\ket{J_L,m_L}$ where $J_L = N_L/2$ and $m_L = (N_{L,\uparrow} - N_{L,\downarrow})/2$.
The initial state of the logic ions can be written as,
\begin{equation}
\ket{\psi}_L = \sum_{m_L = - J_L}^{J_L} c_m\ket{J_L, m_L}.
\end{equation}
The amplitude is given by~\cite{arecchi1972}
\begin{equation}
c_m = \binom{2J_L}{m_L+J_L}^{1/2}\left(\frac{1}{\sqrt{2}}\right)^{2J_L},
\end{equation}
where $\binom{N}{k}$ is the binomial coefficient~\cite{arecchi1972}.
Each state in the basis $\ket{J_L,m_L}$ is displaced by $|\alpha| = 2m_L \alpha_0$ during the opening interferometer pulse and subsequently by $|\alpha| = -2 m_L \alpha_0$ during the closing interferometer pulse. 
When thinking about the case of a single spectrosocpy ion,
the full interferometer sequence produces the state,
\begin{equation}
\ket{\psi} = \sum_{m_L = - J_L}^{J_L} c_{m_L}e^{4im_L\alpha_0\beta_0\sin(\phi_M)}\ket{J_L,m_L}\ket{\beta}_M\ds,
\end{equation}
where each term $\ket{J_L,m_L}$ traces a unique path in phase space and acquires a geometrical phase of $4im_L \alpha_0 \beta_0 \sin(\phi_M)$.  
Since $m_L = (N_{L,\uparrow} - N_{L,\downarrow})/2$, each phase factor $e^{4im_L \alpha_0 \beta_0 \sin(\phi_M)}$ contains $N_{L,\uparrow}$ factors of $e^{2i \alpha_0 \beta_0 \sin(\phi_M)}$ and $N_{L,\downarrow}$ factors of $e^{-2i \alpha_0 \beta_0 \sin(\phi_M)}$.  
Therefore, at the end of the interferometric sequence the logic ion state can be factorized as,
\begin{equation}
\ket{\psi}_L = \left[\frac{1}{\sqrt{2}}\big(e^{-2i \alpha_0 \beta_0 \sin(\phi_M)}\dl + e^{2i\alpha_0\beta_0 \sin(\phi_M)}\ul\big)\right]^{\otimes N_L}.
\end{equation}
Each logic ion thus acts as an independent sensor of the spectroscopy ion displacement amplitude.

\section{Detection efficiency with multiple spectroscopy ions}

In considering multiple spectroscopy ions we make similar assumptions regarding the phase and amplitude of the state-dependent forces as we did for the logic ions.  
Namely, the forces are applied in-phase with equal coupling to each ion.  
The condition that controls the magnitude of the force is the internal state of the spectroscopy ion (i.e., $\ketda_S$, $\ketua_S$ and $\ket{c}_S$ with corresponding force amplitudes $F_{S, \downarrow} = F_S$, $F_{S, \uparrow} = -F_S$ and $F_{S, a} = 0$, respectively. 
Relative to the phase of the initial displacement on the logic ions, the phase of the spectroscopy ion displacement is set to $\pi/2$, such that the ion wavepackets trace out rectangles in phase space.
  
A typical spectroscopy sequence consists of three operations: 
(1) A series of pulses on both the logic ions and the spectroscopy ions to prepare them to a desired initial state. 
(2) An interrogation pulse (or series of pulses) on the spectroscopy ions to excite the $\ketda_S\leftrightarrow\ket{c}_S$ transition.
We assume this operation has no coupling to the ions' shared motion such that the state of the spectroscopy system at the end of the interrogation operation can be written as $\ket{\psi}_S = (c_{S, \downarrow}\ketda_S + c_{S,c}\ket{c}_S)^{\otimes N_S}$.
(3) An interferometric sequence to determine the number of ions $N_{S,\downarrow}$ remaining in the ground state after the spectroscopy operation.  
After this operation, any coherence between the states $\ketda_S$ and $\ket{c}_S$ should have no effect on the outcome of the measurement. 
We are interested only in the populations $P_{S,\downarrow} = |c_{S,\downarrow}|^2$ and $P_{S,a}= |c_{S,a}|^2 = 1-|c_{S,\downarrow}|^2$.  We ensure that this is the case by 
interacting with the measurement procedure only through the $\ketda_S$ state which has no coupling to $\ket{c}_S$ during this operation.

After the spectroscopy sequence, state detection proceeds through the same interferometer sequence described for a single spectroscopy ion above.  
In the second step of the interferometer, a resonant force applied to the spectroscopy ions results in displacement $D(\beta)$ where $\beta = iN_{S,\downarrow}\beta_0$. The total phase accumulated between logic ion states $\ketda_L$ and $\ketua_L$ is proportional to $N_{S,\downarrow}$,
\begin{equation}
 \theta = 2N_{S,\downarrow}\alpha_0\beta_0 \sin \phi_M.
\end{equation}
    
The choice of $4N_{S}\alpha_0\beta_0 \sin \phi_M = \pi$ bounds the geometric phase to $\theta \in [0, \pi/2]$.  If we define the collective spin basis states of the spectroscopy ion system $|J_S, m_S\rangle$ similarly to the case of the logic ions, the full state at the end of the detection sequence can be written as,
\begin{equation}\label{eq:psi_tot}
    \ket{\psi} = \sum_{m_S = -J_S}^{J_S} c_{m_S}(\Phi)
    \left[ 
        {\rm cos}\left(\frac{\pi}{4}\theta_S\right)\ketda_L 
         + 
        i{\rm sin}\left(\frac{\pi}{4}\theta_S \right)\ketua_L \right]^{\otimes N_L} 
    \ket{(J_S-m_S)\beta_0}_M\ket{J_S,m_S},
\end{equation}
where $\theta_S = 2 (J_S-m_S)/J_S$, $c_{m_S}(\Phi)$ is the amplitude of the state and  $\ket{J_S, m_S}$ as a function of the atom-laser phase at the end of a Ramsey sequence $\Phi = (\omega_L-\omega_A)T_R$. Here,
\begin{equation}
    c_{m_S}(\Phi) = \sqrt{\binom{2J_S}{m_S+J_S}}\sin^{J_S+m_S}\left(\frac{\Phi}{2}\right)\cos^{J_S-m_S}\left(\frac{\Phi}{2}\right), 
\end{equation}
This results in all possible states of the spectroscopy ion system  mapping to a unique transition probability in the logic ion system, $P_{L, \uparrow} = 1 - P_{L, \downarrow} =  {\rm sin}^2(\pi N_{S,\downarrow}/2 N_S)$.

To determine the projection noise limit, we consider the signal-to-noise ratio for Ramsey spectroscopy as a function of the Ramsey phase $\Phi = \delta T_R$.  The uncertainty in a phase measurement is given by,
\begin{equation}
\Delta\Phi = \frac{\Delta m_L}{|d\langle m_L\rangle/d\Phi|},
\end{equation}
where $(\Delta m_L)^2$ is the variance of a $\hat{J}_{L,z}$ measurement on the LIs, and $d\langle m_L\rangle/d\Phi$ is the slope of its expectation value as a function of the Ramsey phase.  
In general, projection noise of the logic ion system will add to projection noise of the spectroscopy ion system to set a fundamental limit to the signal-to-noise ratio achievable using this readout method.  
However, given a sufficient number of logic ions or, equivalently, a sufficient number of repetitions of the readout sequence, the added projection noise from the logic ion system can be made negligible.  
In that case, we recover the standard quantum limit $\Delta\Phi = 1/\sqrt{N_S}$.

\begin{figure}
    \centering
    \includegraphics[width=12.7cm]{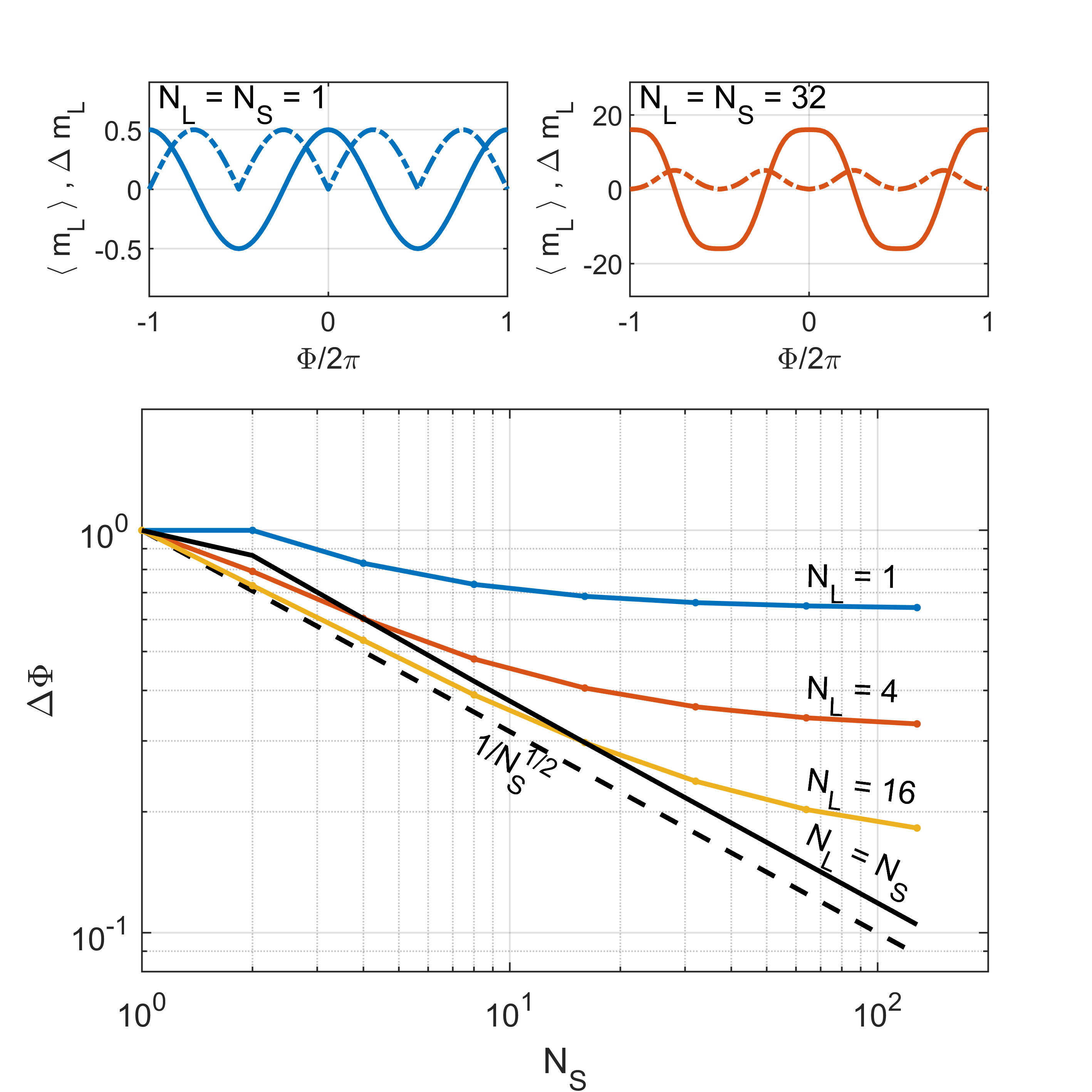}
    \caption{\label{fig:SNR}
    Projection noise limit for Ramsey phase measurements using $N_S$ SIs and $N_L$ LIs based on simulation. In all cases, we assume a single repetition of the multi-ion readout sequence.  The upper figures show expectation values (solid lines) and standard deviations (dashed lines) of $m_L$ measurements as a function of the Ramsey phase for two examples: $N_L = N_S = 1$ (upper left) and $N_L = N_S = 32$ (upper right).  The projection noise as a function of $N_S$ is given in the lower plot for several values of $N_L$. The standard quantum limit (dashed, black line) is determined only by $N_S$, but there is added projection noise from the logic ion subsystem.  We find the choice of $N_L = N_S$ results in a small increase in the projection noise, above the standard quantum limit, which can be further suppressed with increasing $N_L$, or equivalently by using multiple detection cycles.}  
\end{figure}

In Fig.~S1 we numerically compute the projection noise limit for different cases of $N_S$ and $N_L$ assuming the optimum Ramsey phase $\Phi = \pi/2$.  
For $N_S = 1$, independent of $N_L$, the detection sequence produces a maximally entangled state of the two ions such that there is no additional projection noise from the logic ion.  
For $N_S > 1$ added projection noise from the logic ion system causes a deviation from the standard quantum limit.  
This added noise is suppressed with increasing $N_L$, or equivalently by repeated detections as described in the main text.  
One feature to note is that for $N_S>2$ the expectation value $\langle m_L\rangle$ does not vary sinusoidally as a function of the Ramsey phase. This higher error signal slope $|d\langle J_{L,z}\rangle/d\Phi|$ at $\theta = \pi/2$ increases sensitivity to laser phase fluctuations and partially compensates for the additional projection noise from the logic ion system.  
      
\section{Numerical model beyond the Lamb-Dicke limit}

The geometric phases acquired through a Schr\"{o}dinger cat interferometer as described here are first-order insensitive to the initial motional state,
making this detection method applicable to cases outside the ground state or 
even approaching the Doppler cooling limit.
However, we have thus far assumed that the ions remain in the Lamb-Dicke regime during the entire 
experimental sequence.
This imposes a significant constraint in our experiments and, as we show below,
violation of this assumption explains most of the observed loss of contrast in $P_{\downarrow, L}$ as a function of $|\beta|$.

The state-dependent forces required for the logic and spectroscopy systems are typically produced using laser beams that have significant field gradients relative to the size of the ion motional state.
Such field gradients can be realized in multiple ways, which are usually broken into two categories.
The first one acts on the $\sg_z$ logic basis (referred to as a $\sg_z$-type force), and the second one acts on the orthogonal rotated basis $\sg_\phi$ (referred to as a $\sg_\phi$-type force).
The $\sg_z$-type force requires preparing 
$\ket{\psi}_z = \ketda_{S, z} \left(\ketda_{L,z} + \ketua_{L,z} \right)/\sqrt{2}$ at the beginning, 
while the $\sg_\phi$-type requires 
$\ket{\psi}_\phi 
    = \ket{\rightarrow}_{S} \left(\ket{\rightarrow}_{L} + \ket{\leftarrow}_{L} \right)/\sqrt{2}
    = \ketda_{L, z} \left(\ketda_{S, z} + \ketua_{S, z} \right)/\sqrt{2} $ state.
The rotation from $\ketda_z$ to $\ket{\rightarrow}$ is done by employing a $\pi/2$ pulse on the qubit transition of the corresponding ion system.
In this paper, we focus on the realization of a $\sg_\phi$-type force using a bichromatic laser field. 
We note that it is possible to realize our scheme based on a $\sg_z$-type force as well.
In the following, we will drop the subscript $z$ or $\phi$ without causing ambiguity to show the versatility.

The Hamiltonian that describing a two-level atomic ion trapped in a harmonic potential along the $\hat{z}$-direction can be expressed as
\begin{equation}
    \Ham_0 = \frac{\hbar \omega_0}{2} \sg_z + \hbar \omega_M (\am \amd + \frac{1}{2}),
\end{equation}
where $\omega_0$ is the resonant frequency between the qubit $\ketda_z$ and $\ketua_z$ states,
$\am$ and $\amd$ are ladder operators of the harmonic oscillator.
The interaction between this ion and a single mode laser can be expressed as~\cite{wineland1998}
\begin{equation}
    \Ham_I = \frac{\hbar\Omega}{2} (\sg_+ + \sg_-) [
        e^{i(\bm{k}\hat{z}-\omega_L t+\phi_L)}
    + h.c.].
\end{equation}
where $\bm{k}=2\pi/\lambda=\omega_L/c$ is the wave vector and $\phi_L$ is the phase of the laser.
The Rabi frequency, $\Omega$, describes the coupling strength between the laser field and the ion's internal states.
If the frequency of the laser $\omega_L$ is set close to resonance ($\delta = \omega_L-\omega_0 \ll \Omega$), it is convenient to transform to the interaction picture using 
$\Ham_{\text{int}} = \hat{\mathcal{U}}^\dagger\Ham_I\hat{\mathcal{U}}$
where $\hat{\mathcal{U}}=\textrm{exp}[-(i/\hbar)\Ham_0 t]$. 
The interaction Hamiltonian becomes:
\begin{equation}
    \label{eq:interaction}
    \Ham_{\text{int}} = \frac{\hbar\Omega}{2} [
        \sg_+ e^{i\eta (
            \am e^{-i\omega_M t} + \amd e^{i\omega_M t}
        )}e^{-i(\delta t-\phi_L)}
    + h.c.]
\end{equation}
We introduce $\eta=\bm{k} \cdot \bm{z}_0$ as the Lamb-Dicke parameter, where $\bm{z}_0$ is the extent of the ground state wave function in direction $\bm{z}$.
If we strictly consider the Lamb-Dicke limit, $(2\langle n \rangle +1)\eta^2 \ll 1$, where $\langle n \rangle$ is the mean occupation number of the motional mode,
we can simplify Eq.~(15) by expanding it to the lowest order of $\eta$:
\begin{equation}
    \Ham_{\text{int}} = \frac{\hbar\Omega}{2} [
        \sg_+ (1+i\eta (\am e^{-i\omega_M t} + \amd e^{i\omega_M t})
              e^{-i(\delta t-\phi_L)}
    + h.c.]
\end{equation}
If $\delta = 0$, the laser field (named a ``carrier pulse'') would oscillate ions between $\ket{\downarrow}_z$ and $\ket{\uparrow}_z$ states.
After neglecting the rapidly varying terms proportional to $\textrm{exp}(\pm i(\omega_L + \omega_0)t)$ terms (rotating-wave approximation, RWA), the Hamiltonian has the form:
\begin{equation}
    \Ham_{\text{car}} = \frac{\hbar\Omega}{2} 
    (\sigma_+ e^{i\phi_L} + \sigma_- e^{-i\phi_L}).
\end{equation}

Similar to a single mode laser, we write the Hamiltonian of a bichromatic laser fields with frequency components
$\omega_1 = \omega_0 - \omega_M + \delta_s$ (the red detuned component) and 
$\omega_2 = \omega_0 + \omega_M + \delta_s$ (the blue detuned component) interacting with the ion simultaneously.
Both components are detuned by $\delta_s$ from the motional sideband, 
where $\delta_s$ is the detuning of the mean frequency of the bichromatic field from the SI qubit resonance.
Using the Lamb-Dicke approximation and RWA, the Hamiltonian becomes
\begin{equation}
    \begin{split}
     & H_{int} = H_{rsb} + H_{bsb}, \\
     & H_{rsb} = \frac{\eta \Omega}{2} (a\sigma_+e^{i(\delta t+\phi_r)} + a^\dagger \sigma_- e^{-i(\delta t+\phi_r)}), \\
     & H_{bsb} = \frac{\eta \Omega}{2} (a^\dagger \sigma_+e^{i(\delta t + \phi_b)} + a \sigma_-e^{-i(\delta t + \phi_b)}).
    \end{split}
\end{equation}
If we define $\phi_s = (\phi_b+\phi_r)/2$ as the qubit phase and
$\phi_m = (\phi_b-\phi_r)/2$ as the motional phase, we can introduce a rotated basis
$\hat{\sigma}_\phi = e^{-i\delta_s t}e^{-i \phi_s} \hat{\sigma}_+ + e^{i\delta_s t}e^{i\phi_s} \hat{\sigma}_-$.
Note that $\sg_\phi =\sg_x$ for the case $\delta_s=0, \phi_s=0$
The interaction Hamiltonian now becomes:
\begin{equation}
    \Ham = \frac{\eta\Omega}{2}\sg_\phi (\am e^{i\phi_M} +\amd e^{-i\phi_M}).
\end{equation}
When the detuning of both the red and blue components are set on resonance with the motional sideband ($\delta_s=0$),
the unitary evolution of the Hamiltonian implements the state-dependent displacement operator:
\begin{equation}
    \hat{\mathcal{D}}(\alpha(t)) = e^{\sg_\phi [\alpha(t)\amd - \alpha(t)^*\am]},
\end{equation} 
where $\alpha(t) = -i\eta\Omega t e^{-i\phi_M}/2$. 

When the condition $(2\langle n \rangle +1)\eta^2 \ll 1$ is not satisfied, higher order terms in the Taylor expansion of $e^{i\eta(\am+\amd)}$ must be included.
This is done by replacing $\am$ and $\amd$ with the matrix $\bra{n}e^{i\eta(\am+\amd)}\ket{n-1}$ and $\bra{n-1}e^{i\eta(a+a\dagger)}\ket{n}$ respectively. 
We have\cite{wineland1998}
\begin{equation}
    \bra{m}e^{i\eta(a+a\dagger)}\ket{n} = e^{i\eta^2/2} \eta^{|m-n|} L_n^{|m-n|}(\eta^2) \left(\frac{n!}{m!}\right)^{1/2},
\end{equation}
where $n, m$ are Fock states included in the simulation and $L_n^\alpha$ is the generalized Laguerre polynomial
\begin{equation}
    L_n^\alpha (X) = \sum_{m=0}^n (-1)^m
    \begin{pmatrix}
    n+\alpha \\
    n-m \\
    \end{pmatrix}
    \frac{X^m}{m!}.
\end{equation}

\begin{figure}
    \centering
    \includegraphics[width=12.0cm]{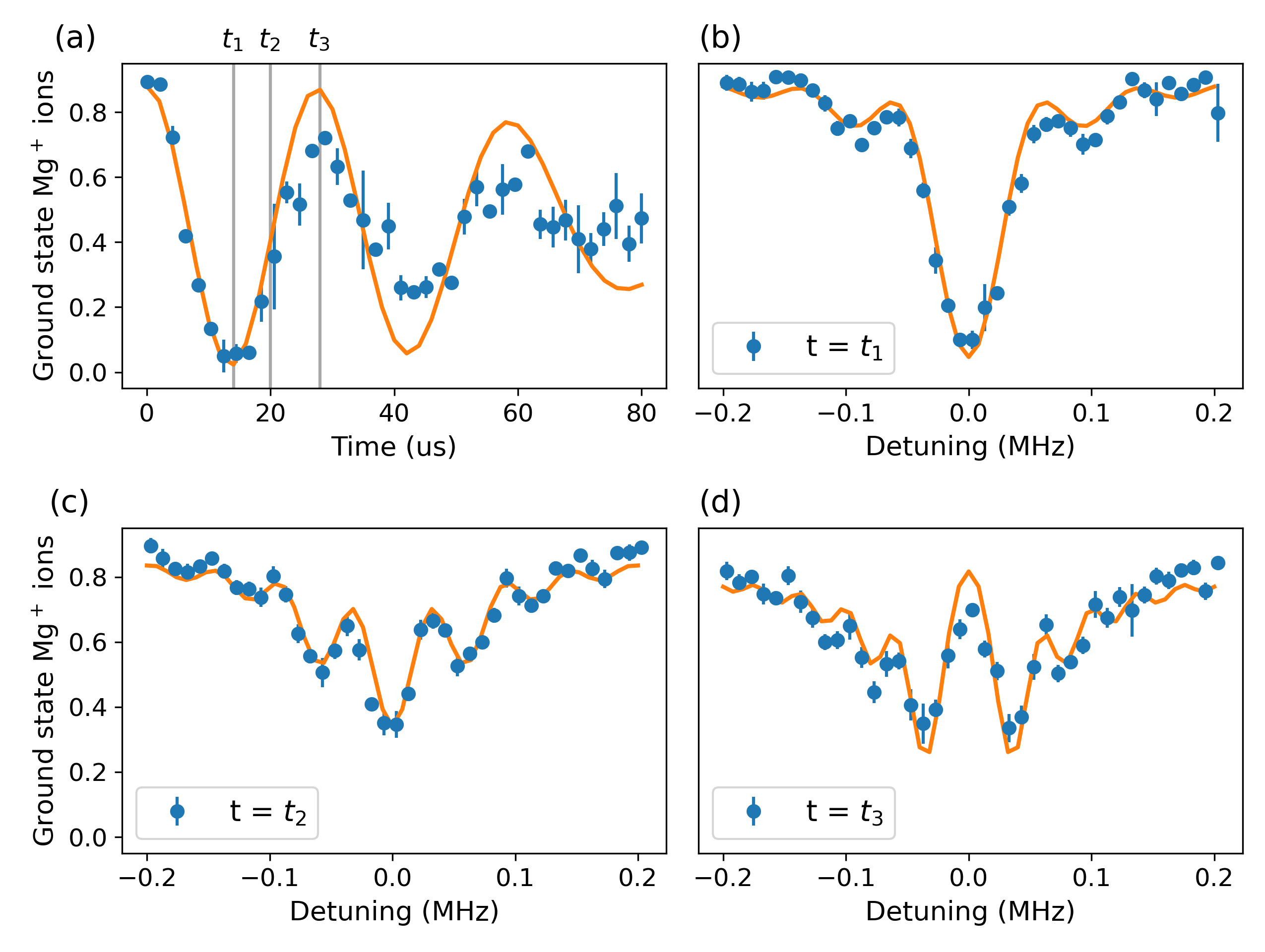}
    \caption{\label{fig:freq_scan}
        Comparison the numerical simulation with experimental data.
        All blue points represent experimental data taken with a pair of $\Mg$-$\Al$ ions after ground state cooling of the COM mode.
        All orange traces are the results of numerical simulation using the same parameters calibrated from the experiments.
        (a) We first scan the duration of the driving force applied on the spectroscopy ion and manually adjust the $\eta_S$ and $\eta_L$ about 10\% to make sure the period of the simulated result is the same with the experimental result. Then, we use the same parameters to run the simulation at $t=t_1$ (a), $t=t_2$ (b) and $t=t_3$ (c). Note that the data used in (a) is the same one shown in Fig.1(b) in main text.
        An additional loss of contrast ($24\%$) is added to all four simulation results.
    }
\end{figure}

We solve the following master equation pulse-by-pulse to get the evolution of the density matrix $\rho(t)$ of the system consisting of LIs and SIs:
\begin{equation}
    \dot{\rho}(t) = -\frac{i}{\hbar} [\Ham_{int}(t), \rho(t)].
\end{equation}
Initially, the ions are prepared in $\ket{\psi} = \ketda_L (\ketda_S + \ketua_S)/\sqrt{2} $ state.
There are three laser pulses involved in our simulation:
(1) a LI-SDD pulse on the LIs with $\phi_r=\pi/2, \phi_b=-\pi/2$;
(2) a SI-SD pulse on the SIs with $\phi_r=\phi_b=0$; 
and (3) a LI-SDD$^{-1}$ pulse on the LIs with $\phi_r=3\pi/2, \phi_b=\pi/2$.
As can be seen, our simulation requires knowledge of the Rabi frequency $\Omega$ of each laser and the Lamb-Dicke parameters, $\eta$, of LIs and SIs, respectively.
We measure $\Omega$ of the SI-SDD and the LI-SDD by observing the Rabi oscillation of the qubit transitions of $\Mg$ and $\Al$ with a pair of $\Mg$-$\Al$ ions.
The Lamb-Dicke parameter is given by the ratio of the Rabi frequency of the first sideband and carrier transition after cooling to the motional ground state: $\eta=\Omega_{sb}/\Omega_{carrier}$.

In Fig.~S2, we compared experimental data with our numerical simulation results.
The experiments here are done on a pair of $\Mg$-$\Al$ ions after ground-state cooling.
The simulation results for $P_{\downarrow, L}(t)$ is first compared with the experimental data shown in Fig.~S2(a) where $t$ is the duration of the SI-SDD pulse.
We observe that the measured Lamb-Dicke parameters for the LIs $\eta_L$ and SIs ($\eta_S$) have a relative uncertainty of 10\%,
so we manually adjust both $\eta_S$ and $\eta_L$ a bit to make sure the Rabi frequency of the simulation result agrees with the experimental data.
We also find that there is a loss of contrast (24\%) in our experiments.
The reason is not clear, so we simply reduce the contrast on the simulation results.
Then, we use exactly the same parameters to simulate $P_{\downarrow, L}(\delta_s)$ at different pulse times: $t_1= 14~\mu$s, $t_2= 28~\mu$s and $t_3= 14~\mu$s.
Note that the result at $t_2= 14~\mu$s (Fig.~S2(b)) shows the same data as Fig.~2(a) in the main text.
The rotation of the spin basis $\sg_\phi$ creates complicated fringes when $\delta_s \neq 0$, which is explained by our model.

Since the Lamb-Dicke approximation requires $(2\langle n \rangle +1)\eta^2 \ll 1$,
it is straightforward to see that the non-zero temperature of the ions will lead to stronger coupling with higher-order terms.
We use a pair of LI and SI to simulate the effect of this coupling.
To simplify the discussion, we assume $\eta_L=\eta_S=\eta=0.14$, $\phi_M=\pi/2$, and $N_S=N_L=N$.
We define the contrast as $1-P_{\downarrow,L}(\theta)$ where we set $\theta=\pi$ to flip the state of the LIs.
Both ions are cooled to a thermal distribution of the motional mode, whose temperature is described by $\langle n \rangle$. 
As we expect, the contrast decreases as the temperature increases. 
In Fig.~S3(a), We simulate and find that the loss of contrast is reduced as we add more ions, which reduces $\eta$ by a factor of $\sqrt{N}$, 
where $N$ is the total number of ions.

\begin{figure}
    \centering
    \includegraphics[width=12.0cm]{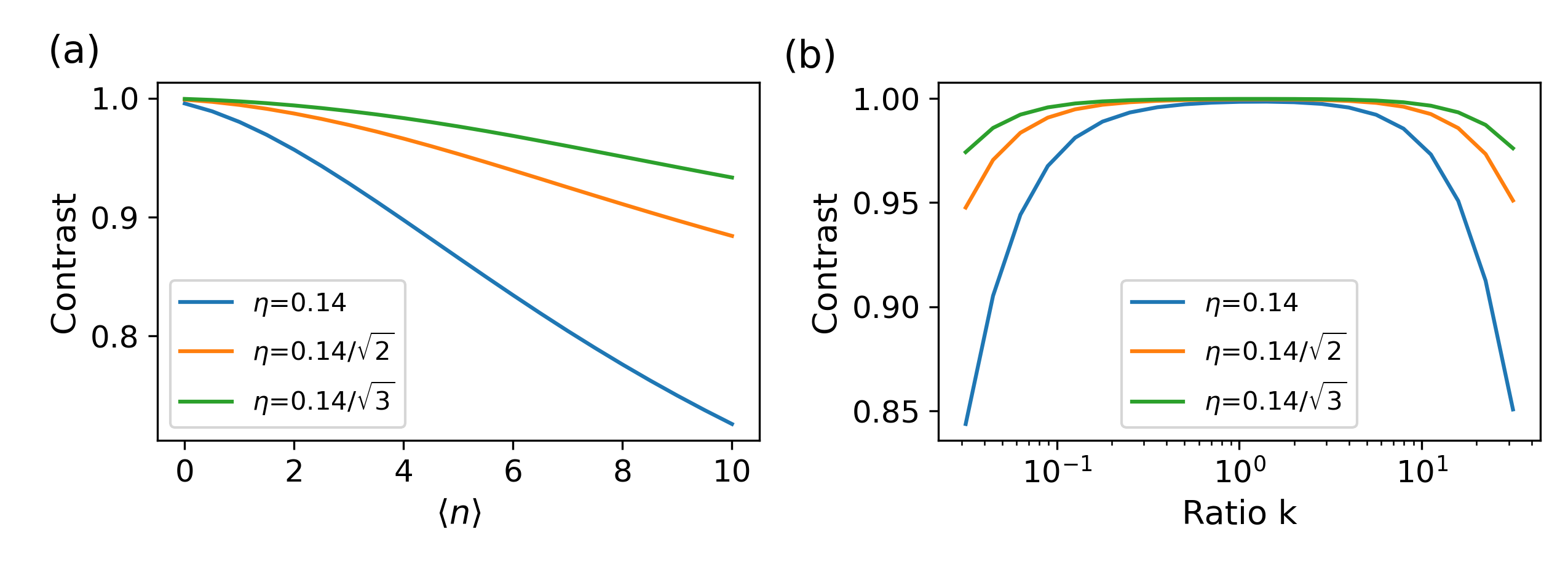}
    \caption{\label{fig:contrast}
        Loss of contrast due to coupling beyond the Lamb-Dicke limit. 
        We assumed $\eta_L=\eta_S=\eta$, $ N_L = N_S = N$ in this simulation.
        $\langle n \rangle$ indicates the mean occupation number of ion's thermal distribution of the motional mode. $k=|\alpha|/|\beta|$ is defined as the ratio of displacement on LI and SI. Although the contrast keeps decreasing when the ions move further from the Lamb-Dicke limit, reducing the Lamb-Dicke parameters (experimentally, by trapping more ions) suppressed this effect significantly.  
    }
\end{figure}

Although one may freely choose $|\alpha|$ and $|\beta|$ subject to the constraint $4|\alpha||\beta|=\pi$, it is wise to consider the maximum displacement in phase space $\delta_{max} = |\alpha|+|\beta|$. 
The mean motional occupation number at this point is $\langle n \rangle = (|\alpha| + |\beta|\rm{cos}\phi_M)^2 + |\beta|^2$ should be minimized to avoid the coupling beyond the Lamb-Dicke limit. 
When we define $k=|\alpha|/|\beta|$ and assume $\phi_M=\pi/2$, the
simulation suggest a maximum contrast (Fig.~S3(b)) when $k=1$.
Once again, this source of reducing contrast can be suppressed by trapping more ions.

\section{Experimental realization of the spin-dependent displacement}

It is important to maintain the phase stability between the red and blue components of the bichromatic laser field.
This is done through a radio-frequency (RF) mixing scheme as shown in Fig.~S4.
The 280 nm (for the $\Mg$ logic ions) and 266.9 nm (for the $\Al$ spectroscopy ions) laser beams are frequency and phase shifted using a series of acousto-optic modulators (AOMs).
These AOMs are driven by direct digital synthesizers (DDSs) to control the frequency detuning, intensity, and phase of the lasers.

In the case of the 267 nm laser system, we use a double-passed AOM to provide a frequency tuning range of 50 MHz. 
Then, another single passed AOM centered at $\omega_{0, \textrm{Al}} = 180$ MHz  is used for switching the beam.
A DDS is employed to generate a RF signal at the motional frequency of the COM mode $\omega_M$,
which is mixed together with a second source at $\omega_{0, \textrm{Al}}$ to drive the bichromatic laser field.
When the 267 nm laser is tuned to be on resonance with the $\sAl \leftrightarrow \pAl$ transition of $\Al$ ions, 
the mixed-signal $\omega_{0, \textrm{Al}} \pm \omega_m$ creates a bichromatic laser field consisting of two frequency components that are equally spaced from resonance.
The mixing depth is selected that only the first-order sidebands are used to drive the motional sidebands.

\begin{figure}
    \centering
    \includegraphics[width=12.0cm]{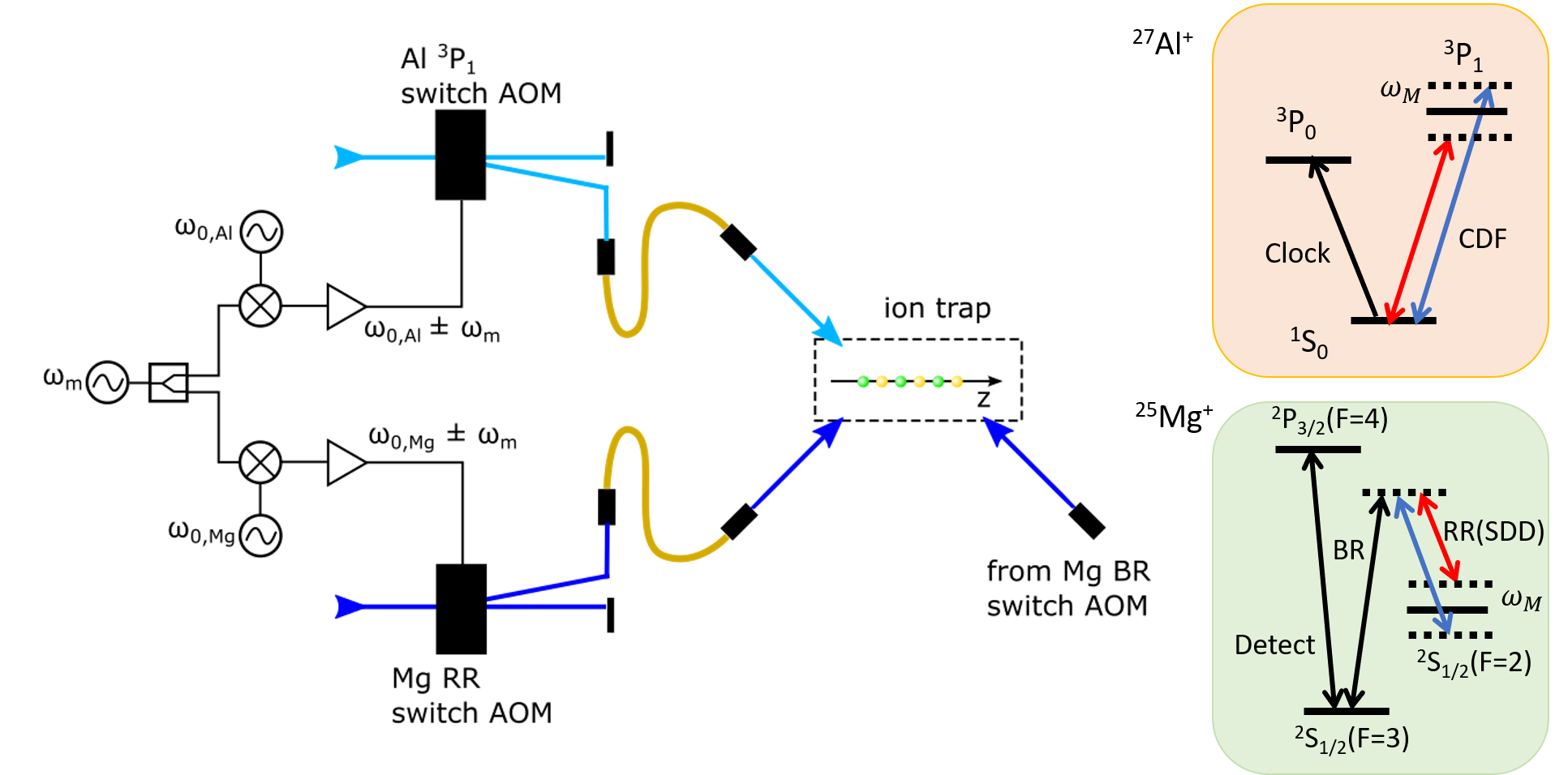}
    \caption{\label{fig:result_qnd}
    RF mixing scheme for the bichromatic laser field. The motionoal frequency $\omega_M$ is COM mode of the ions. $\omega_{0, \textrm{Al}}$ and $\omega_{0, \textrm{Mg}}$ are calibrated to drive the carrier transition on $\Al$ and $\Mg$ ions, respectively (see the energy levels on the right panel).
    }
\end{figure}

The 280 nm laser system used for driving Raman transition on $\Mg$ uses two double-pass AOMs to split the laser into two parts, with a frequency difference of $\omega_{\textrm{HF}}=1.2$ GHz.
Both the red Raman (RR) and blue Raman (BR) components go through additional single-pass AOMs centered at $\omega_{0,\textrm{Mg}}=$ 300 MHz to provide switching of the beams.
Similar to the single-pass AOM in the 267 nm laser beam line, an RF signal with a frequency of $\omega_M$ from the same DDS is mixed with $\omega_{0,\textrm{Mg}}$ to generate the bichromatic laser field in the RR beam.
This design makes both the LI-SDD and the SI-SDD insensitive to the motional phases of the ions~\cite{haljan2005}, 
which will allow for scaling to larger ion crystals since the motional phases vary between different positions in the ion crystal.


Both $\omega_{0, \textrm{Mg}}$ and $\omega_{0, \textrm{Al}}$ are calibrated to drive the carrier transition of $\Mg$ ions and $\Al$ ions, respectively.
The laser intensity of both the 266.9 nm laser and the 280 nm laser are controlled using separate calibration pulses at the beginning of the experimental sequence.
Feedback control is applied to the drive amplitudes of the AOMs to stabilize the intensity of the laser beams.
We noticed that the motional frequency drifts in a range of about 200 Hz during the experiments, so an interleaved lock-in method is used to track $\omega_M$ roughly every minute.

\bibliography{multi_ions}